\documentclass[usenatbib,iop,apj,tighten,numberedappendix,twocolappendix]{aeb_emulateapj_2015}
\usepackage[usenames,dvipsnames]{color}
\definecolor{darkblue}{rgb}{0.0,0.0,0.3}
\usepackage{bm}
\usepackage{graphicx}
\usepackage{amsmath}
\usepackage{amssymb}
\graphicspath{{images/}}
\usepackage{color}

\usepackage[normalem]{ulem}

\newcommand{\comment}[1]{}

\def\IMF{{\rm IMF}}

\def\kms{{\rm km\,s^{-1}}}

\begin{document}

\title{
Resolving Microlensing Events with Triggered VLBI
}

\author{
  Mansour~Karami\altaffilmark{1,2}, 
  Avery~E.~Broderick\altaffilmark{1,2},
  Sohrab~Rahvar\altaffilmark{4,1},
  Mark~Reid\altaffilmark{3},
}
\altaffiltext{1}{Perimeter Institute for Theoretical Physics, 31 Caroline Street North, Waterloo, ON N2L 2Y5, Canada}
\altaffiltext{2}{Department of Physics and Astronomy, University of Waterloo, 200 University Avenue West, Waterloo, ON N2L 3G1, Canada}
\altaffiltext{3}{Harvard-Smithsonian Center for Astrophysics, 60 Garden Street, Cambridge, MA 02138, USA}
\altaffiltext{4}{Sharif University of Technology, Azadi Ave. Tehran, Iran PO Box: 11365-11155}  

\shorttitle{Resolving Microlensing Events with Triggered VLBI}
\shortauthors{Karami et al.}

\begin{abstract}
Microlensing events provide a unique capacity to study the stellar remnant population of the Galaxy.  
Optical microlensing suffers from a near complete degeneracy between the mass, the velocity and the distance.
However, a subpopulation of lensed stars, Mira variable stars, are also radio bright, exhibiting strong SiO masers.  These are sufficiently bright and compact to permit direct imaging using existing very long baseline interferometers such as the Very Long Baseline Array (VLBA).  We show that these events are relatively common, occurring at a rate of $\approx 2~{\rm yr^{-1}}$ of which $0.1~{\rm yr^{-1}}$ are associated with Galactic black holes.  Features in the associated images, e.g., the Einstein ring, are sufficiently well resolved to fully reconstruct the lens properties, enabling the measurement of mass, distance, and tangential velocity of the lensing object to a precision better than 15\%. Future radio microlensing surveys conducted with upcoming radio telescopes combined with modest improvements in the VLBA could increase the rate of Galactic black hole events to roughly 10~${\rm yr}^{-1}$, sufficient to double the number of known stellar mass black holes in a couple years, and permitting the construction of distribution functions of stellar mass black hole properties.
\end{abstract}


\section{Introduction} 

Gravitational lensing presents one of a handful of current
observational windows on the dark universe.  The distortion of
background galaxies by galaxy clusters has provided a striking
demonstration of general relativity \citep{2010CQGra..27w3001B}. More importantly, it has
enabled the unique reconstruction of the projected mass density of the
cluster dark matter haloes \citep{refregier,pac}.  While the dynamics of stars and gas
probe the dark matter halo within $10^2~{\rm kpc}$ of the central
galaxy \citep{sal}, gravitational lensing currently provides the only
means to detect and study the halo on larger scales.  By combining
such measurements, it has become possible to map the dark matter
distribution in the local universe \citep{shan}.

The detailed structure of a gravitationally lensed image may be
separated into the intrinsic structure of the lensed object and a
distortion imposed by the lens that depends upon the lens location and
mass distribution.  A typical scale is set by the Einstein angle,
this is exactly the angular scale of the Einstein ring cast by a point lens,
\begin{equation} 
\theta_E 
=
\sqrt{\frac{4GM (D_S-D_L)}{c^2 D_S D_L} }\,,
\label{thetae}
\end{equation}
where $M$ is the total source mass and $D_S$ and $D_L$ are the source
and lens distance, respectively.  For galaxy clusters this can span
many arcminutes.

Gravitational lensing within the Galaxy also occurs
when the angular separation of a background star and a foreground object becomes comparable 
to $\theta_E$.  This differs
from the gravitational lensing of clusters and galaxies in two
important respects.  Firstly, the angular size of the lensed image has
a typical scale of $\theta_E\approx 1 (M/M_\odot)^{1/2}~{\rm mas}$,
well below the resolution of existing optical telescopes.  As a
result, direct optical imaging has not been feasible.  Nevertheless,
the conservation of brightness, a consequence of Liouville's theorem,
implies substantial magnifications
of the total flux associated with
the magnified image, typically increasing by more than an order of
magnitude.  Secondly, the peculiar motions of objects within the
Galaxy cause the source-lens system to evolve on time-scales comparable
to 
\begin{equation}
t_E \equiv \frac{\theta_E}{\left|\bm{\mu}_S - \bm{\mu}_L\right|}\,,
\label{eq:tE}
\end{equation}
where $\bm{\mu}_S$ and $\bm{\mu}_L$ are the source and lens apparent
angular velocities, respectively.  For typical values in the Galaxy
$t_E$ ranges from days to months, depending on the source
mass and distance of lens and source from the observer.  The relationship between $t_E$ and the lens structure implies a
characteristic light curve, referred to as a microlensing event.

The rarity of serendipitous lens-source alignments require the monitoring of large numbers of potential sources to identify candidate microlensing events.  For this purpose optical surveys of dense star fields have been undertaken by a number of groups, both to constrain the contribution of stellar remnants to the Galactic dark matter budget \citep{moniez2010} and more recently to find extrasolar planets which results in short time-scale features in the microlensing light curve \citep{gaudi2012}.  These have successfully excluded remnants with masses between $10^{-7}M_\odot$--$10M_\odot$ as a candidate for dark matter within the Galactic halo \citep{lasserre,alcock2001}.

Motivated by the potential to detect extra-solar planets, recent years
have seen the initiation of high-cadence, large-area surveys.  These
have been enabled by the development of large CCD arrays for
telescopes with wide fields of view.  As a result, the 
OGLE\footnote{http://ogle.astrouw.edu.pl/ogle4/ews/ews.html} (Optical Gravitational Lensing Experiment)
collaboration recorded  $\approx2\times10^3$
events in 2013, using the Early Warning System \citep{earlyw} to alert the follow-up 
telescopes for monitoring microlensing events with better time-coverage
. The Korean Microlensing Telescope Network (KMTNet)
will increasing the number of events to $6000$ per year with a 10
minute cadence operating continuously \citep{kmtnet}.

Reconstruction of microlensing events usually only determines
the $t_E$.  As a direct consequence, the mass of the lens, and the
distances and velocities of the lens and source suffer from
a fundamental degeneracy.  Methods for breaking the mass-distance
degeneracy typically requires higher-order effects.  If the
microlensing parallax can be measured -- an 
asymmetry in the light curve induced by the non-uniform motion of
the Earth -- the lens distance and the mass of the lens can be constrained 
\citep{gould92,rah03}. 
Parallax effect can also be employed by monitoring microlensing events from earth and space
simultaneously. Spitzer has been used to this end, measuring microlensing parallax and thus 
providing an additional constrain \citep{2015arXiv150807027S}.
Alternatively the source finite-size effect could be used to break the degeneracy. If
the angular impact parameter is smaller than the source angular size the point source
approximation is no longer valid and the finite-size effect becomes important.
By analyzing the light curve an additional relation between angular source size and angular Einstein 
radius could be found.
In both cases, the lens and the source apparent
motions must be subsequently measured after the event to remove these
from consideration \citep[see, e.g.,][]{2014JKAS...47..215G}, which is only possible for
stellar lenses and explicitly excludes those associated with stellar
remnants.

Directly resolving the image would greatly simplify the
reconstruction of the lens parameters.  The typical scales of
microlensing events are well matched to those achievable by very long
baseline interferometry (VLBI) at centimetre wavelengths, e.g., with the Very
Long Baseline Array (VLBA).  Unfortunately, imaging microlenses with the VLBA
requires compact radio bright sources with typical brightness
temperatures in excess of $10^{10}~{\rm K}$, well above the typical
stellar temperatures.  However, the unprecedented rate at which
microlensing events are being identified by optical surveys makes it
possible to leverage rare source properties for this purpose.

Mira variables provide a natural target for optically triggered radio
imaging microlensing experiments.\footnote{\citet{2011MNRAS.411.1780H} explore a number of other potential radio-bright microlensing targets, including the continuum emission from a wide class of giants.  Of these, only the SiO masers from Miras and active galactic nuclei have sufficiently high brightness temperatures to be amenable to VLBI imaging.}  Miras are asymptotic giant branch (AGB)
stars with month time-scale pulsations largely determined by convection
in their envelopes \citep{wood2006convection}.  While the optical luminosities of Miras can
vary over three orders of magnitude \citep{2002ApJ...568..931R}, the typical peak luminosities are
on the order of $10^3 L_\odot$, ensuring that they are among the
brightest objects in microlensing survey samples \citep{Soszynski13}.
With typical masses similar to $1M_\odot$ Miras are relatively numerous despite having AGB lifetimes of roughly a million years.
Given the number of stellar targets, optical depth of lenses, and typical lensing timescales the rate of microlensing events with hundred percent detection efficiency is given by
\begin{equation}
\frac{N_{events}}{N_{stars}T_{obs}} = \frac{2}{\pi}\frac{\tau}{<t_E>}.
\label{eq:Nevents}
\end{equation}
Assuming there are $4\times10^4$ Mira stars in the bulge (see Section \ref{sec:Mira}), the optical depth of microlensing events toward the Galactic bulge is $\tau\simeq 2.35\times 10^{-6}$ \citep{sumi}, and a typical time scale of one month for the microlensing events (i.e. $<t_E>\simeq 1$ month) we therefore expect to observe roughly $1$ Mira-source microlensing events per year.

Of particular importance here is the presence of SiO masers in the extended
atmospheres of Miras.   Typical sizes of the masing regions are $1$~AU,
corresponding to angular sizes of $0.1$~mas for distances characteristic of bulge stars \citep{2007ApJ...659..378R}.  Because of the
nonthermal nature of the maser process, the spots can have brightness
temperatures as high as $5\times10^{10}~{\rm K}$, well above the thresholds
for imaging with the VLBA \citep{2007ApJ...659..378R}. 
Within the Galactic center, the inner parsec of
the Galactic bulge, the SiO masers associated with evolved stars have
been imaged with the VLBA at angular resolutions of 0.7~mas,
limited by the interstellar electron scattering in that region
\citep{2003ApJ...587..208R,2007ApJ...659..378R}.  Outside of the Galactic center, it should be possible to
image these with resolutions approaching 0.3~mas.

Here we explore the rates of suitable microlensing events anticipated
by current and future surveys and the precision with which the
associated lens parameters can be reconstructed.  We propose using
optical microlensing surveys to trigger follow-up VLBA observations
based on source location in the color-magnitude diagram, anticipated
event duration, and the lack of a bright lens counterpart.
A subset of these will be Miras, previously identified via the observation of long-time instrinsic variability.
Massive lenses are more heavily represented in long-duration events,
and those without obvious stellar counterparts are more likely to be
associated with massive compact objects.

In section \ref{sec:stats} we describe a Monte Carlo simulation of microlensing events, discuss the relative rates at which Miras are expected to be lensed by various objects, including stellar remnants, and how event selection can be optimized for imaging based on optical properties.
Section \ref{sec:rates} discusses the absolute rates of radio-bright black hole microlensing events of a variety of potential surveys.
Section \ref{sec:radio} describes the resulting images for a simple maser model and estimates the accuracy with which the lens properties can be reconstructed.
The implications for constructing mass and distribution functions for the Galactic remnant population is discussed in \ref{sec:disc}.
Finally, conclusions are collected in \ref{sec:C}.

\section{Stellar Remnants in Simulated Microlensing Surveys} \label{sec:stats}

The rates of microlensing events associated with compact remnants depends on the distribution of sources and lenses within the Galaxy and the detection efficiency of current and upcoming microlensing surveys.  Here we describe simulations of an OGLE-like microlensing survey and estimate the number of microlensing events we anticipate to be associated with the various Galactic remnant populations.  To do this, we perform mock surveys assuming the monitored stellar field is located in the Galactic bulge.

Necessary inputs are the mass, velocity, color, and magnitude distributions of potential sources and the mass and velocity distributions of potential lenses. For simulation of microlensing events, we adapt the detection efficiency in terms of the Einstein crossing time of OGLE survey sources \citep{ogleiii}. We describe the details of the model here.

\subsection{Galactic Distribution Model and Stellar Lens/Source Population Model}
To generate mock microlensing events we require the spatial and velocity distributions of the stellar targets and lenses.

For the structure of the Galaxy, we use thin disk model and standard bulge model from \citet{bt}.  The density distribution in disk is modeled in cylindrical coordinates by a double exponential function,
\begin{equation}
\rho_{D}(R,z) = \frac{\Sigma}{2H} \exp
\left(\frac{-(R-R_{\odot})}{R_d} \right) \exp \left( \frac{-|z|}{H}
\right) \ ,
\end{equation}
where $\Sigma$ is the column density of the disk at the Sun position, $H$ the height scale and $R_d$ the length scale of the disk. The distribution of the lens transverse velocity with respect to the line of sight is established from
solar motion and the local lens velocity distributions (see Table \ref{tabmodel}).

The bar is described in a Cartesian frame positioned at the galactic center with the major axis $x$ tilted by $\phi= 45^\circ$ with respect to the Galactic center-Sun line. The bar density is given by
\begin{equation}
\begin{gathered}
\rho_{B} = \frac{M_{B}}{6.57 \pi abc} e^{-r_s^{2}/2}\\
r_s^{4} =
\left[ \left( \frac{x}{a} \right)^{2} + \left( \frac{y}{b} \right)^{2}
  \right]^{2} + \frac{z^{4}}{c^{4}}\,,
\end{gathered}
\end{equation}
where $M_{B}$ is the bulge mass, and $a$, $b$ and $c$ are the scale length factors. The dispersion velocity of bulge stars is $110~{\rm km~s^{-1}}$. For the density of the spheroid structure we take the following function \citep{robin03},
\begin{equation}
\rho_{spher} = 
\begin{cases}
\rho_0 (a_c/8.5 \mathrm{kpc})^{-2.44} & a \leq a_c\\
\rho_0  (a/8.5 \mathrm{kpc})^{-2.44} & a \ge a_c \
\end{cases}
\end{equation}
$a^2 = R^2 + z^2/(0.76)^2$, $a_c$, $\rho_c$ and spheroid velocity dispersion are given in table (\ref{tabmodel}).

\begin{deluxetable}{c l l c c}
\tablecaption{Assumed parameters of the galactic model, The model is partly adopted from \citep{rah03}.\label{tabmodel}}
\startdata
\hline
\hline 
Structure   & \multicolumn{2}{c}{Parameter}
&\multicolumn{2}{c}{Value} \\ \hline
    & \multicolumn{2}{l}{$\Sigma\ (M_{\odot}\ {\rm pc}^{-2})$} & \multicolumn{2}{c}{50} \\
    & \multicolumn{2}{l}{$H\ ({\rm kpc})$}                 & \multicolumn{2}{c}{0.325} \\
Disk    & \multicolumn{2}{l}{$R_d\ ({\rm kpc})$}               & \multicolumn{2}{c}{3.5}  \\
   
\cline{2-5}
    &velocity   &$\sigma_{r}~(\kms)$       & \multicolumn{2}{c}{34.} \\
    &disper-    &$\sigma_{\theta}~(\kms)$  & \multicolumn{2}{c}{28.} \\
    & sions     &$\sigma_{z}~(\kms)$   & \multicolumn{2}{c}{20.} \\
\hline

    & \multicolumn{2}{l}{$M_{B}\ (M_{\odot})$} & \multicolumn{2}{c}{$1.7\times 10^{10}$}    \\
    & \multicolumn{2}{l}{$a\ ({\rm kpc})$} &   \multicolumn{2}{c}{1.49} \\
Bulge  & \multicolumn{2}{l}{$b\ ({\rm kpc})$} &   \multicolumn{2}{c}{0.58} \\
 & \multicolumn{2}{l}{$c\ ({\rm kpc})$} &   \multicolumn{2}{c}{0.40} \\
\cline{2-5}
     & velocity &$\sigma~(\kms)$        &  \multicolumn{2}{c}{110}  \\ \hline

    & \multicolumn{2}{l}{$\rho_{0}\ (M_{\odot} {\rm pc}^{-3})$} & \multicolumn{2}{c}{$0.932\times 10^{-5}$}    \\
Spheroid    & \multicolumn{2}{l}{$a_{c}\ ({\rm kpc})$} &   \multicolumn{2}{c}{0.5} \\
\cline{2-5}
     & velocity &$\sigma~(\kms)$        &  \multicolumn{2}{c}{120}  \\ \hline

    & \multicolumn{2}{l}{$\rho_{H\odot}\ (M_{\odot} {\rm pc}^{-3})$} & \multicolumn{2}{c}{0.008}    \\
   & \multicolumn{2}{l}{$R_{c}\ ({\rm kpc})$} &   \multicolumn{2}{c}{5.0} \\
Halo    & \multicolumn{2}{l}{$M\ in\ 60 \ {\rm kpc}\ (10^{10} M_{\odot})$}   &   \multicolumn{2}{c}{51}     \\
\cline{2-5}
     & velocity &$\sigma~(\kms)$        &  \multicolumn{2}{c}{200} \\

\enddata
\end{deluxetable}

Combining the density distribution of Galaxy, kinematics, mass function and stellar population of stars, we can generate microlensing events in terms of $t_E$, $M_l$, $D_l$, $D_s$, $\theta_E$ and $\pi$, where $\pi$ is the parallax parameter that is described in the Appendix. 
We use the detection efficiency of OGLE survey in terms of Einstein crossing time (i.e. $\epsilon(t_E)$) to select the observed microlensing events in our simulation \citep{ogleiii}. A general overview about the simulation of microlensing events can be found in \citet{rahvar2015}.

\subsection{Remnant Population Model}

\begin{figure}
\includegraphics[width=\columnwidth]{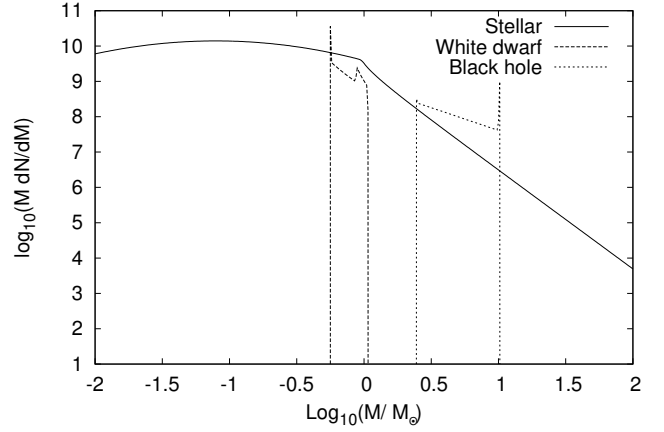}
\caption{Assumed mass functions of stars (solid), white dwarfs (dashed), and black holes (dotted) in the disk and bulge.  The spike features at low (white dwarf) and high (black hole) masses correspond to regions in which the remnant mass is nearly independent of that of the progenitor.  These do not substantially impact the results here.} \label{massf1}
\end{figure}

\begin{figure}
\includegraphics[width=\columnwidth]{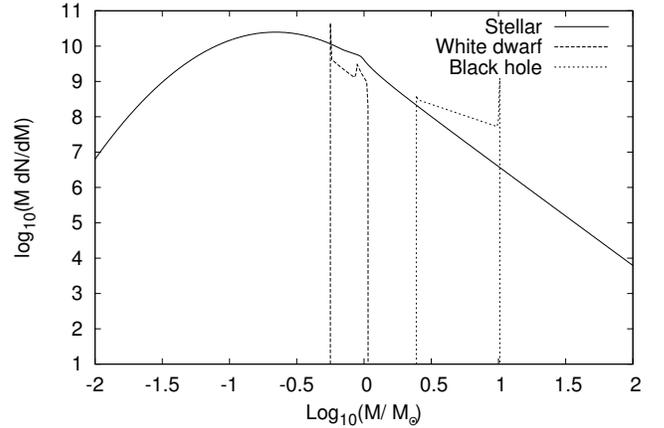}
\caption{Assumed mass function of stars (solid), white dwarfs (dashed), and black holes (dotted) in the Spheroid. The spike features at low (white dwarf) and high (black hole) masses correspond to regions in which the remnant mass is nearly independent of that of the progenitor.  These do not substantially impact the results here.} \label{massf2}
\end{figure}

In addition to the known stellar populations, gravitational lens candidates include the Galactic population of compact objects.  In principle, these may be comprised of the remnants of stellar evolution and primordial objects.  We make the conservative assumption that the latter are absent, and consider only the remnants of massive stars.

The end point of stellar evolution depends primarily on the mass of the progenitor.  We consider two classes of remnants, white dwarfs and black holes.  We ignore the intermediate neutron stars, which we expect to contribute marginally to the high-mass tail of the white dwarf population and otherwise be indistinguishable from them using microlensing observations alone.  Hence in practice the remnants are effectively white dwarfs/neutron stars and black holes.

Encoded in the remnant population is the Galactic high-mass star formation history.  Once the relationship between the zero-age main sequence progenitor mass ($M_{\rm MS}$) and the final remnant mass ($M_R$, where $R$ may be WD or BH) is specificed, the mass function\footnote{We define the mass function to be the number of  objects per logarithmic decade, i.e., $dN/d\log_{10}M$.} ($\phi(M_R)$) is given solely in terms of the star formation rate ($\Lambda(t)$) and initial mass function of main sequence stars ($\Phi(M_{\rm MS})$):
\begin{equation}
\phi(M_R)
=
\int dt \frac{\Lambda(t-T_{\rm MS}) \Phi(M_{\rm MS})}{d\log_{10} M_R/d\log_{10} M_{\rm MS}}\,,
\end{equation}
where $T_{\rm MS}\approx 10 (M_{\rm MS}/M_\odot)^{-1.5}~{\rm Gyr}$ is the lifetime of a main sequence star with mass $M_{\rm MS}$ associated with the remnant mass $M_R$, and the integration is over the entire Galactic history.  Effectively, this is simply the number of remnants formed over the history of Galactic star formation, excluding those stars that remain on the main sequence.  For this we adopt the star formation rate from \citet{Hart04}, which peaks roughly 10~Gyr ago, and the ``Disk and Young Clusters'' stellar initial mass function ($\IMF(M)$) from  \citet{Chab03}, though employing the ``Universal'' IMF from \citet{Krou01} makes no discernible difference.  

We relate $M_{\rm WD}$ and $M_{\rm MS}$ via the piecewise continuous expression presented in \citep{Sala09}, obtained for white dwarfs in open clusters by comparing the white dwarf ages inferred by cooling models and cluster age from isochrone fitting:
\begin{equation}
M_{\rm WD}/M_\odot =
\begin{cases}
0 & m_{\rm MS}\le0.5588\\
0.01 m_{\rm MS} + 0.5418 & 0.5588<m_{\rm MS}\le1.7\\
0.134 m_{\rm MS} + 0.331 & 1.7<m_{\rm MS}\le4\\
0.047 m_{\rm MS} + 0.679 & 4\le m_{\rm MS}<8\\
0 & 8\le m_{\rm MS}\,,\\
\end{cases}
\end{equation}
where $m_{\rm MS}\equiv M_{\rm MS}/M_\odot$.

The relationship between $M_{\rm BH}$ and $M_{\rm MS}$ is highly dependent upon the uncertain evolution of massive stars and the details of core collapse supernovae.  As a result, above $30M_\odot$ it is unclear what the slope of the $M_{\rm BH}$-$M_{\rm MS}$ relationship is \citep[see Figure 9 of][]{Frye12}.  Here we make a moderately conservative assumption that the resulting black hole mass is independent of $M_{\rm MS}$, setting
\begin{equation}
M_{\rm BH}/M_\odot = 
\begin{cases}
0 & m_{\rm MS}\le8\\
0.3 m_{\rm MS} & 8<m_{\rm MS}\le33\\
10   & 33<m_{\rm MS}\,,
\end{cases}
\end{equation}
which lies in the middle of the permitted range presented in \citet{Frye12} and is roughly consistent with narrow black hole mass range inferred from X-ray binaries by \citep{Ozel10}.

The resulting stellar and remnant mass functions for the disk, bulge and spheroid are shown in Figures \ref{massf1}, \ref{massf2}.  The spikes in the low and high mass ends of the white dwarf and black hole mass functions are associated with regions in which the remnant mass are nearly independent of the progenitor mass, and contain a finite number of objects.  For both the disk and the bulge there is a break in the stellar mass function where stars produced during the star formation peak are leaving the main sequence, roughly $1 M_\odot$.

Because neither white dwarfs nor black holes are expected to experience strong kicks at formation, we assume the same velocity dispersion for the remnants as the surrounding stellar population. This is notably not the case for neutron stars, which often experience sufficient kicks to be launched into the halo, diluting their density in the disk and therefore their representation in the lens sample for current and planed microlensing surveys.The observation of pulsars with ages less than $3~\rm{Gyr}$, indicates the pulsar birth velocity of about $400~\kms$ \citep{pulsar}.

\subsection{Mock Observation Statistics}
 In order to simulate the microlensing events, we choose the source stars from a Hipparcos-like color-magnitude distribution \citep{Perryman97}, distributed proportional to the density of matter. After selecting the source star with a given absolute color and magnitude, the position of star in the color--magnitude diagram changes due to the distance modulus and interstellar extinction. Those stars brighter than the limiting magnitude of the microlensing survey (which is taken about 19 in I-band in this simulation) are selected for the observation. It should be noted that the results are not sensitive to the limiting magnitude and 
could be applied to microlensing surveys with different limiting magnitudes. 

We note that not all the microlensing events generated in this simulation are observable in reality. Depending on duration of observation, cadence of photometric data and non-observing nights due to bad weather and technical failures, only a fraction of events as a function of Einstein crossing time, $t_E$, can be observed. We adapt the detection efficiency function (i.e. $\epsilon(t_E)$) for selecting the observed events from OGLE published function \citep{ogleiii}. Detailed description about gravitational microlensing formalism can be found in \citet{rahvar2015}. 
 
Each event is constructed by first choosing a random source with the appropriate properties, subsequently a random lens with the appropriate properties, and finally applying the OGLE detection efficiency.  That is,
\begin{enumerate}
\item Source star selection:
  \begin{enumerate}
  \item A source star position is selected with a probability distribution given by the stellar density and geometric factors, i.e.,
    \begin{equation}
      \frac{dP}{dD_S} \propto \rho(D_S) D_S^2\,,
    \end{equation}
  \item The source star stellar type is randomly chosen according to Hipparcos Color-Magnitude distribution.
  \item The stellar component (bulge, disk, spheroid) is selected with weights equal to their density at $D_S$.
  \item The source star velocity is the sum of the bulk Galactic motion for the chosen stellar component at the chosen source location and a random contribution pulled from a Gaussian distribution with the appropriate dispersion (listed in Table \ref{tabmodel}).
  \end{enumerate}
\item Lens selection:
  \begin{enumerate}
  \item A lens position is selected from the stellar density, modified by the lensing cross section, i.e.,
    \begin{equation}
      \frac{dP}{dD_L} \propto \rho(D_L)\sqrt{\frac{D_L(D_S-D_L)}{D_S^2}}\,.
    \end{equation}
  \item A lens component (bulge, disk, spheroid) is randomly chosen according to the local densities of each at $D_L$.
  \item A lens component type (star, white dwarf, black hole) is randomly chosen according to the local fraction of each for the given component.
  \item A lens mass is selected according to the total mass function of the chosen component modified by the approximate lensing cross section; for a fixed observing time the probability of detecting a lens is $\propto \theta_E \propto M^{1/2}$.
  \item A lens velocity is the sum of the bulk Galactic motion for the chosen component and a random contribution pulled from a Gaussian distribution with the appropriate dispersion (listed in Table \ref{tabmodel}).
  \end{enumerate}
\item Detection efficiency cut: Given the lens mass, source and lens velocities, and source and lens positions $t_E$ is computed, and the event is accepted with a probability set by the OGLE detection efficiency.
\end{enumerate}
The result is a set of lensing events with known intrinsic properties (source and lens masses, distances, velocities, and types) selected using a realistic microlensing survey biases.

\begin{figure}
\includegraphics[width=\columnwidth]{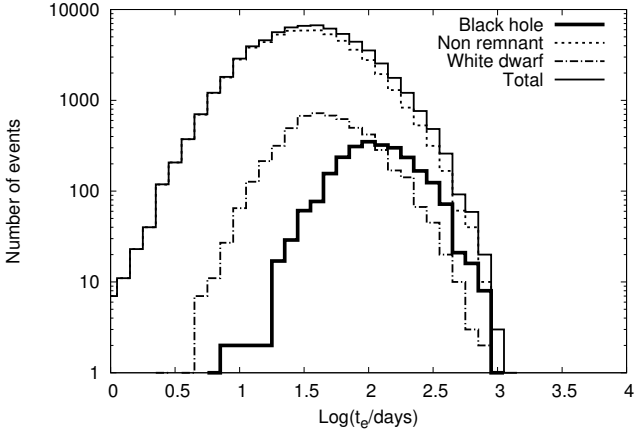}
\caption{Distribution of microlensing events in Einstein 
crossing time from Monte-Carlo simulation. The detection efficiency of OGLE  survey is applied during event selection. The fraction of black hole events to the overall events is $0.035$ and weighted heavily towards long $t_{\rm E}$.}
\label{te_sim}
\end{figure}

Figure \ref{te_sim} shows the distribution in Einstein crossing time of simulated events.  Since in the simulation we flag lenses that are either main sequence or remnants and, we can classify lenses based on lens type.  Here we compare the overall microlensing events with those produced by black holes specifically; the fraction of events with black hole lenses is $0.035$.
This fraction is comparable to the estimate of \citet{gould2000}, which finds 0.01 assuming a 100\% detection efficiency, and thus neglecting the higher detection efficiency of Mira events.

Another important point is the average $\log{(t_E)}$ for the overall events and black hole events are $1.5$ and $2.0$, respectively.  This is unsurprising since the duration of events scales with lens mass as $M^{1/2}$, and is an immediate consequence of the fact that the average black hole mass is roughly an order of magnitude larger than that of the overall sample.  Hence, black hole lensing events are typically more than three times longer than the typical microlensing event, with important consequences for identification and parameter estimation.

\begin{figure}
\includegraphics[width=\columnwidth]{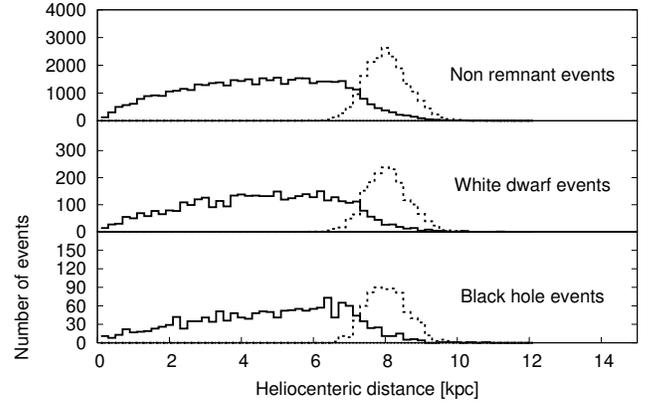}
\caption{Distribution of lens positions in galaxy for various lens populations.  Contributions of the disk (solid) and bulge (dashed) are shown separately; the spheroid contribution is negligible.}
\label{ND_sim}
\end{figure}

Figure \ref{ND_sim} shows the distribution of lens locations in our simulated survey for each of the Galactic components and lens types.  Events are dominated by lenses within the disk (65\%) with the remainder from the bulge (35\%); the spheroid contributes negligibly.  Of note is that the lens distribution within the disk is nearly evenly distributed between the bulge and a heliocentric distance of 2~kpc, implying that microlensing surveys necessarily probe the lens population throughout the intervening disk.  This remains true independent of the lens type.

\section{Radio-Bright Microlensing Event Rates} \label{sec:rates}

Key to the radio imaging of microlensing events is the identification of radio-bright lensed sources.  In the near term this requires the identification of candidate events based on their optical properties alone.  Here we discuss the rates implied by the previous section employing surveys that exploit optical counterparts of radio bright sources (i.e., Mira variables) and the potential rates from future radio microlensing surveys.  This is ultimately dependent on the population and distribution of candidate sources.

\subsection{Miras as Microlensing Sources}

\begin{figure}
\includegraphics[width=\columnwidth]{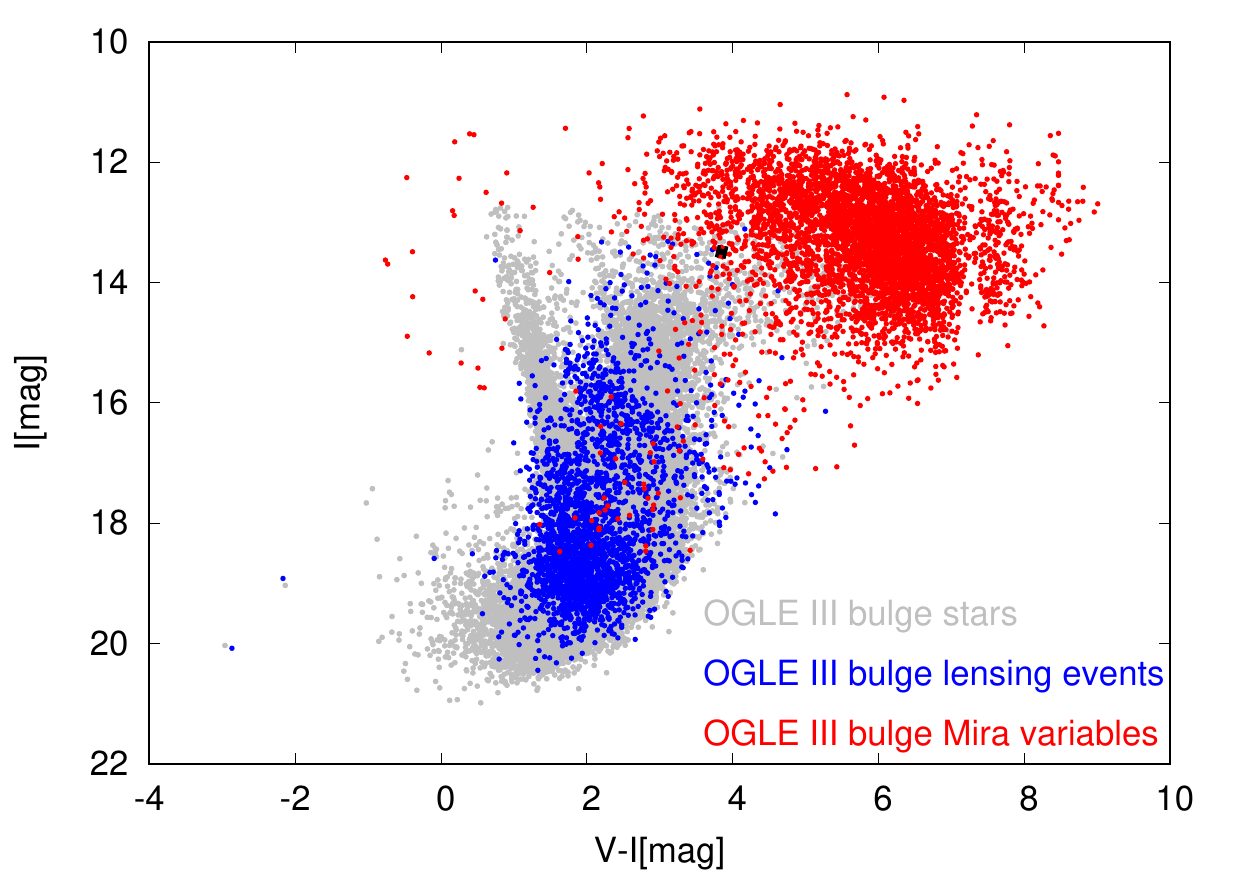}
\caption{OGLE III Color-Magnitude diagram. Grey points are a subsample of the bulge stars monitored by OGLE III. The red and blue points are the bulge Mira variables and Bulge lensing events observed by OGLE III respectively. The small black interval inside the box shows how much a MIII giant will move in HR diagram as a result of blending with a solar type (G2V) lens star located half way between the source and observer.}
\label{goodCMD}
\end{figure}

Miras provide a natural class of optically bright sources with strong radio emission in the form of circumstellar SiO masers.  In most microlensing surveys Miras have already been identified due to their variable nature.  Where they have not been, they may be crudely identified with in the optical color magnitude diagram (CMD), shown for the OGLE III sample in Figure \ref{goodCMD}.  In particular, Miras produce a bright, red cluster that only marginally overlaps with the remainder of the survey targets.
From direct inspection of the OGLE III fields we estimate that as many as 5\% of the objects near the center of the Mira cluster in Figure \ref{goodCMD} are identified Miras.\footnote{ For this purpose we fit the Mira density distribution with a Gaussian and define ``near center'' to be within a single standard deviation from the mean.}  
In Section \ref{goodCMD} we show that there should be nearly 6 times more Miras in the bulge than have been detected thus far, suggesting that it might be possible to increase the fraction of Miras in this region to up to 10\%-15\%, after accounting for the field of view of the OGLE III survey.  
Given their high radio brightness, confirmation that a given candidate source is a Mira can then be quickly obtained by direct radio observation with existing large radio telescopes.
Equally important is that there is little chance for lensing-induced source confusion -- blending of source and lens stars does not move objects appreciably within the region populated by Miras.

Rapid intrinsic stellar variability provides an obvious impediment to the identification and characterization of microlensing events.  For this reason variable stars have been generally identified and excluded from past and ongoing surveys.  Thus, while there are microlensing events reported by OGLE III that penetrate the Mira cluster (see Figure \ref{goodCMD}), these are almost certainly not associated with Miras.  Nevertheless, the presence of such events makes clear that apart from variability there are no intrinsic barriers to including stars lying in the Mira-cluster-region of the CMD in microlensing surveys.  

The variable nature of Miras can be mitigated in a number of ways.  Typical Miras vary with periods of $\approx 1~\mathrm{yr}$, I-band variability amplitudes between $1.4~{\rm mag}$ and $3~{\rm mag}$ \citep{Soszynski13}, and V-band variability amplitudes typically two--three times larger \citep{1990gcvs.book.....K}.  However, black hole microlensing events can be easily distinguished from the intrinsic variability as a result of three important differences.  All of these may be exploited in part due to the high luminosity of Miras, and therefore the high photometric accuracy with which they may be monitored.


First, microlensing light curves have a well-understood structure set by the lens-source geometry and nature of gravitational lensing, characterized by a divergent rise as the source approaches the lens caustics (moderated for the largest magnification events by the finite source size; see Section \ref{sec:genmock} and Appendix \ref{app:lens}).
This differs from the intrinsic variations of Miras.  Thus careful light curve fitting should be able to rapidly distinguish between microlensing and intrinsic variability.  

Second, the amplitude of the flux variation during microlensing events is typically larger than that due to intrinsic variability.  Typical magnifications are of order 200\%, while high-magnification events can exceed an 1000\%.  These correspond to magnitude variations of $0.75~{\rm mag}$ and $2.5~{\rm mag}$, comparable the that caused by intrinsic variations in I-band.  Hence, the microlensing signal is an order unity modification of the underlying source variability in I-band.  At longer wavelengths the intrinsic variability decreases further, on average variations in the magnitude at 1.25~$\mu$m are 20\% of those at optical wavelengths \citep{2002AJ....123..948S}\footnote{ There are instances for which near-infrared variability is nevertheless sizable, e.g., R Cas, where the amplitude of variations at 1.25~$\mu$m are as large as 1.42~mag.  However, the variability continues to drop rapidly with increasing wavelength.}.

Third, the flux variations due to microlensing are achromatic.  Again this is in stark contrast to the intrinsic variability of Miras, which varies dramatically with wavelength. Thus, multi-band monitoring should be readily able to disentangle the two variability components.

We make no further attempt to assess the efficiency with which Mira microlensing events can be identified, i.e., we presume that the intrinsic variability of Miras does not result in a significant inefficiency.  As a result, our rate estimates may be considered to be optimistic.

\subsection{Bulge Miras Microlensing Rates}
\label{sec:Mira}
The OGLE III survey monitored approximately $150$ million stars in Galactic bulge from 2001 to 2009 in search for gravitational microlensing events and provided us with the largest catalogue of microlensing events available.  OGLE used the Early Warning System (EWS) to detect ongoing microlensing events \citep{earlyw}. EWS uses Difference Image Analysis photometry and by analyzing all the stars in the field flags the potential microlensing events \citep{alard}.  In doing so it currently identifies and filters out variable stars. The flagged stars go through several tests including a visual inspection of the light curve and after satisfying all the criteria they are announced as a microlensing candidate.
Figure \ref{goodCMD} shows  the CMD for OGLE III microlensing events \citep{ogleiii}. The
lensing events have not gone through blending correction for consistency reasons since it cannot be done for non lensed stars. 

 
Miras are a few orders of magnitude brighter than M-dwarfs, increasing their representation in microlensing surveys.  An advantage of using bright stars is that blending by background and lens stars do not affect the brightness of the source star significantly (Figure \ref{goodCMD}). Moreover, from the lens equation, and estimating the mass of the lens star, we can calculate the contribution of blending and correct the position of source star in CMD \citep{muraki,miyake}.

Within the OGLEIII catalogue there are currently $6528$ Miras \citep{Soszynski13}.  The larger number of Miras inferred in the bulge suggest that this can be readily enlarged by surveying the entire bulge region.
A rough estimate of the number of Miras in the bulge can be made via the infrared surface brightness.

We follow the approach used in \citet{Matsunaga05}.  The method is based on using the empirical relation  between Mira number density and infrared surface brightness to infer the total number of Miras in the bulge.
\begin{figure}
  \includegraphics[width=\columnwidth]{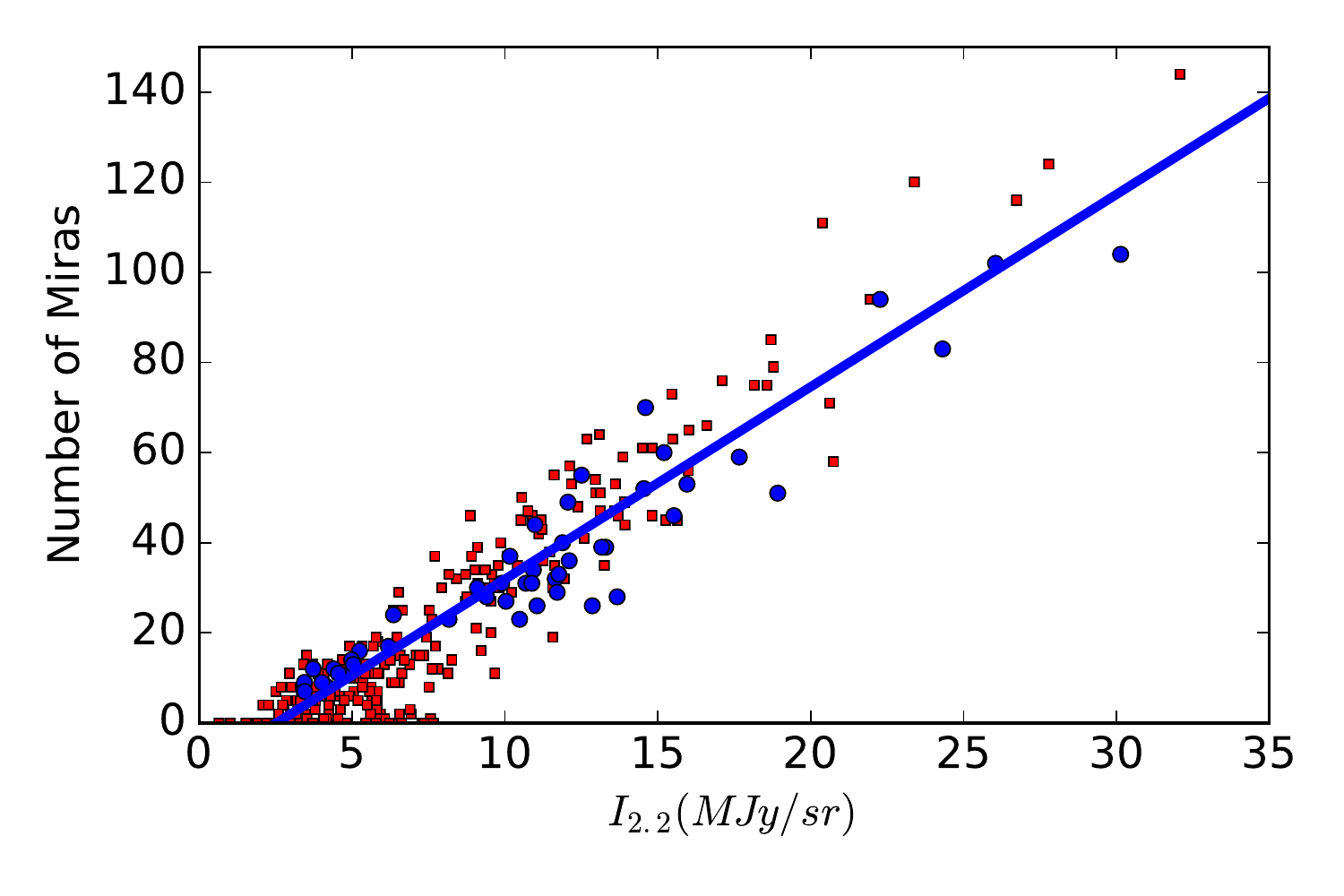}
  \caption{Number of Miras in OGLEII (blue circles) and OGLEIII (red circles) fields vs COBE/DIRBE $2.2 \mu m$ flux in the same regions.  The best linear fit to OGLEII data is shown by the blue line, consistent with \citet{Matsunaga05}.}
  \label{MiraCorrelation}
\end{figure}

We employ the infrared brightness measurements from  Diffuse Infrared Background Experiment (DIRBE) aboard the Cosmic Background Explorer (COBE) satellite. DIRBE made full sky brightness maps in ten infrared bands ranging from $1.25 \mu \rm{m}$ to $240 \mu \rm{m}$. For our purpose we make use of the zodical-light-subtracted infrarred mission-average map at $2.2 \mu\rm{m}$. The maps are available online at \url{http://lambda.gsfc.nasa.gov/product/cobe/dirbe_prod_table.cfm}.

Before using the DIRBE brightness measurements it must be corrected for dust extinction and the contribution from the Galactic disk removed.  Following \citet{Matsunaga05} we correct for dust using an empirical relationship between the K-band extinction $A_K$ and infrared colour:
\begin{equation}
A_K = 0.73 \times (-2.5\log{(I_{1.25}/I_{2.2})}+0.14)\,,
\end{equation}
where $I_{1.25}$ is the $1.25 \mu\rm{m}$ zodiacal-light-subtracted mission-average DIRBE flux.

To remove the Galactic disk contribution we make use of fits to the infrared brightness maps outside the bulge.  \citet{Matsunaga05} adopted an exponential function in Galactic longitude to model the disk
\begin{equation}
  I(l, b) = I(0, b) \exp{(-|l|/l_0(b))}\,,
\end{equation}
where the scale-height in Galactic longitude $l_0(b)$ is a function of Galactic latitude.  This function was fit to the $2.2 \mu\rm{m}$ DIRBE maps at high Galactic latitutdes, $10^\circ<|l|<45^\circ$.  Motivated by the exponential vertical structure of the Galactic disk, we also estimated the disk contribution by fitting an exponential function in Galactic {\em latitude}.  The resulting bulge brightness estimates, and thus the number of bulge Miras, is insensitive to which fitting function is used.

As shown in Figure \ref{MiraCorrelation}, the number of Miras in a given field is strongly correlated with the corrected infrared surface brightness.  Like \citet{Matsunaga05} we find a high-quality linear relationship between the Mira number counts in OGLEII fields and the corrected DIRBE $2.2\mu\rm{m}$.  This remains true when Mira number counts in OGLEIII fields are added.  Linear fits to the latter gives an updated ratio of Mira number density to $2.2\mu\rm{m}$ surface brightness of $4.3 {\rm Sr/MJy}$, similar to that found by \citet{Matsunaga05}.  With a total $2.2\mu\rm{m}$ flux of $0.9{\rm MJy}$ in the bulge ($-10^{\circ}<l<10^{\circ}$ and $-10^{\circ}<b<10^{\circ}$), this implies that there are around $4\times10^4$ bulge Miras.\footnote{This number is insensitive to alternative fitting functions for the Mira number--$2.2\mu\rm{m}$ surface brightness relation.  We tried linear fits with non-zero offsets, corresponding to an unsubtracted background component, and non-linear fits, none of which substantially changed this number.}

\subsection{Current and Near-future Optical Survey Microlensing Rates}
At least two improved microlensing surveys are underway, exceeding OGLE III in three ways: higher cadence, deeper magnitude limits, and increased sky coverage.  OGLE IV, in operation since 2010, has a bulge field of $\approx 330~ \mathrm{deg}^2$, 3.3 times larger than that of OGLE III.  The resulting OGLE IV EWS event rate is three times larger than that of OGLE III; OGLE III EWS detected about $650$ events in 2008 whereas OGLE IV EWS detected almost $2000$ events in 2013 and over $2000$ in 2014.  The forthcoming The Korean Microlensing Telescope Network (KMTNet) consists of three dedicated $1.6~\mathrm{m}$ telescopes located in Australia, South Africa, and South America, each with $4~ \mathrm{deg}^2$ field of view.  Thus, KMTNet anticipates detecting $\approx 2300$ microlensing events per year with constant cadence of almost 10~min \citep{kmt2014}.

The high luminosity of Miras and the long time-scales of black hole microlensing events imply that the first two, higher cadence and deeper magnitude limits, will at best produce modest improvements in the Mira microlensing rate.  High cadences will vastly over-sample black hole microlensing light curves, and thus while important for planet searches are unlikely to be helpful for the study of Galactic remnants.

Even within the extincted region near the Galactic center Miras remain visible due to their high intrinsic luminosity.  Since optically dim stars do not produce strong radio emission (Stellar photospheres of dwarf stars are essentially
undetectable with current radio telescopes), expanding the survey sample to less luminous objects provides little value for the study of black holes.  Nevertheless, the deeper magnitude limits of future surveys will permit the inclusion of regions of high extinction, i.e., dark areas of the bulge, in survey fields \citep{extinction}, marginally increasing the number Miras that can be monitored.  This may be ameliorated by operating microlensing surveys in the infrared as well.

Much more important is the expanded sky coverage.  We have already assumed that the entire Galactic bulge will be monitored in our previous rate estimate.  Further increasing the number of monitored Miras substantially in an optical survey necessarily requires expanding the surveys beyond the bulge.  Moderate modifications to the observing strategies of existing optical surveys present an obvious way in which radio-bright microlensing event rates can be substantially increased.  Because Miras are intrinsically luminous, they lie well above the detection threshold of all ongoing microlensing surveys located anywhere within the Galaxy.  Therefore, the relative paucity of Miras does suggest an alternative strategy: broad and shallow instead of narrow and deep.

Including the Galactic disk increases the number of stellar targets by a factor of $\approx 6$-$7$.  Assuming the stellar population does not vary widely between the bulge and the disk this would produce an commensurate growth in the Mira sample.  Hence, we expect $\approx 2-3 \times 10^5$ Miras in total within Milky way.  Unfortunately, the microlensing optical depth in the disk is roughly a third of that of the bulge as a result of the former's smaller stellar density.
Therefore, after taking into account the disparity in optical depth, the net increase of Mira microlensing events is reduced to a factor of 2-3.  As a result, we estimate that
a survey that monitors the entire Galactic plane to an I-band magnitude limit of 15 would capture radio-bright microlensing events at a rate of $2~{\rm yr^{-1}}$, corresponding to a radio-bright black hole microlensing events at a rate of $0.1~{\rm yr^{-1}}$.

Importantly, these survey strategies need not be exclusive.  Since the objects of primary interest for radio imaging, black holes, result in long-duration microlensing events, even sparse time sampling is sufficient (on the order of once per week).  The OGLE~IV survey has already adopted such a strategy, monitoring the entire Galactic disk with a cadence of 1-2 days and the bulge with a cadence $\lesssim1$~hr \citep{2015AcA....65....1U}.

\subsection{Far-future Optical Microlensing Surveys}
The LSST (Large Synoptic Survey Telescope) is a 8.4 meter telescope equipped with a wide field camera with 9.6 square degrees field of view. It will provide a deep survey of southern sky (over 20000 square degrees) giving 1000 exposures of each patch during 10 years of observations. The optimal cadence of the survey for different areas of the sky is yet to be determined.  The single exposure depth in r-filter will be $\approx 24.5~ \rm{mag}$ and it can yet go deeper ($\approx 26.5~ \rm{mag}$) by co-adding the images.   Being a ground base survey the angular resolution is limited by seeing (0''.7) and ultimately by its pixel size (0''.2) \citep{2008arXiv0805.2366I}.

The mean cadence of $3-4~ \rm{days}$ is much shorter than the typical time-scales of the long duration black hole lensing events. Furthermore the deep survey combined with the fact that Mira variables are bright stars guarantees that most of the Mira lensing events in the Galaxy, LMC, and SMC will be detected by LSST.  Thus, for finding black hole lensing events amenable to radio imaging, the LSST presents a similar capability to current generation microlensing surveys operated with modified observation strategy.

\subsection{A Future Radio-Continuum  Microlensing Survey}
Were a microlensing survey performed in the radio directly the need to identify optically-bright radio counterparts would be eliminated altogether, resulting in the finding of events that are amenable to radio imaging with a near 100\% efficiency. Moreover such a radio survey would permit monitoring the much more numerous compact continuum radio sources. With the advent of a number of rapid-survey radio telescopes (e.g., the Square Kilometre Array \citealt{taylor07}, Canadian Hydrogen Intensity Mapping Experiment \citealt{2014SPIE.9145E..22B}, etc.), enabling rapid nearly all-sky surveys, high-cadence radio transient searches will become common.  A radio microlensing survey is a natural byproduct of these.

The stringent requirements on source size ($\lesssim 1~{\rm mas}$) restricts the potential radio microlensing survey source targets to stars (e.g., Miras) or primarily extragalactic objects.  The details of the lens parameter estimation are only weakly dependent on the distance to the radio source; placing sources at extragalactic distances decreases the Einstein angle by at most 30\%.  

More important is the over-all number of sources, which directly translates into the predicted rate enhancement.  These, in turn, depend on the flux limit and wavelength of interest, both set by the VLBA.  For a $10\,M_\odot$ black hole lens multiple image components are easily visible by the VLBA for frequencies above 3~GHz, providing a natural upper limit of 10~cm on the observation wavelength.  For 10~min integration times with 500~MHz bandwidths the VLBA flux limit should be near 20~$\mu$Jy and 40~$\mu$Jy at 22~GHz and 43~GHz, respectively; where required we adopt a flux limits of 30~$\mu$Jy.\footnote{We assume system-equivalent flux densities are roughly 500~Jy and 1000~Jy at 22~GHz and 43~GHz, respectively.}  Large-amplification events will reduce the effective flux limit further.  As we will see below, either improvements in the sensitivity of the VLBA or new radio-bright source classes will be necessary to leverage a radio microlensing survey.

Radio-bright active galactic nuclei (AGN) provide a natural class of sources, being both numerous, and more importantly compact.  Conveniently, they are also typically flat-spectrum radio sources, meaning that neither the source identification nor the subsequent microlensing survey need be performed at the same wavelengths at which the events are ultimately imaged.  Thus, to estimate the number of potential sources above the VLBA flux limit we employ the NRAO VLA Sky Survey at 1.4~GHz reported in \citet{1998AJ....115.1693C} \citep[though see also][]{1984ApJ...287..461C,2012ApJ...758...23C} finding roughly $1.6\times10^7$ and $1.5\times10^8$ radio-bright AGN in ellipticals and spirals, respectively, with fluxes above above 30~$\mu$Jy.

These are distributed isotropically, and thus a large fraction of AGN will be visible at high Galactic latitude with correspondingly smaller lensing optical depth. As a result, the average lensing optical depth for an AGN survey to that for the optical surveys that monitor the Galactic bulge alone is approximately $0.01$.  Thus, despite having nearly 290 times the number of targets, a radio microlensing survey of objects above 30~$\mu$Jy would produce a black hole microlensing event rate 1~${\rm yr}^{-1}$, about the same as that from Mira-based optical surveys.  

To obtain an event rate of 10~${\rm yr}^{-1}$, doubling the number of know black holes in two years, requires a flux limit of 3~$\mu$Jy.  This is would require corresponding improvements in the VLBA.  However, in principle this may be achieved via an extended integration time combined with an expanded bandwidth.  Currently, 4~GHz bandwidths supported by 16~Gbps recorders are planned stations participating in millimetre-wavelength VLBI observations \citep{2012arXiv1210.5961W}, and could be deployed to VLBA stations for use at centimetre wavelengths.  Combined with 2~hr integration times, these reach the 3~$\mu$Jy flux limit needed.  Below 3~$\mu$Jy the number of sources scale approximately as $N_{>S}\propto S^{-0.5}$, and thus further growth in the number of sources is a slow function of flux limit.

\section{Radio VLBI Event Reconstruction}\label{sec:radio}

Motivated by the prospect of substantial number of optical microlensing events amenable to radio imaging, we now present illustrative examples of what these may look like for typical event parameters, and discuss the precision with which the event parameters may be reconstructed.

\subsection{Models of Resolved Masers}

SiO masers at $\approx43$~GHz ($\approx7$~mm) have routinely been detected
around late-type giants in the central parsec of the Galaxy \citep{2003ApJ...587..208R}.
Located within the extended atmospheres of Miras, these are typically
within 8~AU of their parent, corresponding to an astrometric offset
of 1~mas in the Galactic center.  Note that this is comparable to the
Einstein angle, $\theta_E$, and therefore the lensing of the maser
emission is distinct from that of optical emission of the star due to the 
finite size effect of source star \citep{Witt}.

The masing spots are resolved both because of their intrinsic structure and as a
result of an interstellar scattering screen that scatter-broadens
images of the Galactic center \citep{1978ApJ...222L...9B,2006ApJ...648L.127B,gwinn2014discovery}.  Because
of the latter effect, even point sources exhibit an extended source
structure with a full-width half-max (FWHM) of $\approx0.7$~mas, effectively
limiting longest baselines that can be employed by the VLBA.  However,
this scattering is highly localized, occurring only for sources in the
immediate vicinity of the Galactic center, and unlikely to limit the
resolution attainable by efforts to image Miras throughout the
Galactic bulge.  

Typical intrinsic sizes for nearby SiO individual maser spots are
1~AU, corresponding to $\approx0.1$~mas at $\approx8$~kpc distance, and therefore also unlikely
to significantly limit the resolution of VLBA observations.  
We ignore the impact of intervening scatter broadening and model the
intrinsic emission from an individual spot by a Gaussian intensity
profile with a FWHM of $0.1$~mas:
\begin{equation}
I_{\rm int}(\bm{\beta}) = I_0 e^{-\left|\bm{\beta}\right|^2/2\sigma_{\rm spot}^2}\,,
\end{equation}
where $\sigma_{\rm spot}= 0.1~{\rm mas}/\sqrt{8\ln2} = 0.042~{\rm mas}$.

In practice, SiO masers from Mira variables often form arc-like structures that are dominated by a handful of individual masing spots.  For black hole lenses the angular sizes of these structures are typically smaller than $\theta_E$, and thus multiple spots are likely to be strongly lensed simultaneously.  For less massive lenses the ring angular angular scale and $\theta_E$ may be more similar.  Nevertheless, because the emission between maser spots remains incoherent\footnote{This is distinct from, e.g., for pulsars, which produce many {\em coherent} spots due to scintillation within the interstellar medium, and therefore require careful consideration of wave-optics effects \citep{2014MNRAS.440L..36P}.}. the resulting lensed image is a linear superposition of a number of individual spots.   Thus, here we consider the simpler problem of a single masing spot to assess the size of the constraints that can be placed on the lens in principle.

Net velocity offsets between the star and masing spots are systematically incorporated into the radio source velocities and thus do not present an additional systematic uncertainty.  However, the masing spots can also evolve in size and luminosity.  Both are unlikely to produce substantial complications since it doesn't impact the separation of the lense images (see Section \ref{sec:pe}).  While the latter can complicate the determination of the radio light curve, this may be ameliorated via the optical light curve.

\subsection{Generating Mock Images} \label{sec:genmock}
Time sequences of mock images are generated via a multi-step process:
beginning with the specification of the positions of the source, lens,
and observer, the mapping of the source to the image plane via the
thin-lens equation, and convolution with a realistic beam.

Over the duration of a lensing event, the position of the source and
lens are assumed to evolve with a fixed velocity:
\begin{equation}
\bm{x}_S = \bm{x}_{S,0} + \bm{v}_S t
\quad\text{and}\quad
\bm{x}_L = \bm{x}_{L,0} + \bm{v}_L t\,,
\end{equation}
respectively.  In practice only the transverse motion is important.  
In contrast, the observer position, i.e., that of the Earth, is
orbital, and thus includes the orbital acceleration:
\begin{equation}
\bm{x}_\oplus = \bm{x}_{\oplus,0} + \int dt \, \bm{\Omega}_\oplus\times\bm{r}_\oplus\,.
\end{equation}
The lens geometry is then fully defined by the distances
\begin{equation}
D_L\equiv \left| \bm{x}_L - \bm{x}_\oplus \right|
\quad\text{and}\quad
D_S\equiv \left| \bm{x}_S - \bm{x}_\oplus \right|\,,
\end{equation}
and the transverse angular displacement
\begin{equation}
\bm{\beta} \equiv 
\frac{(\bm{x}_S-\bm{x}_\oplus)}{D_S}
-
\frac{(\bm{x}_S-\bm{x}_\oplus)}{D_S}\cdot\frac{(\bm{x}_L-\bm{x}_\oplus)}{D_L}
\frac{(\bm{x}_L-\bm{x}_\oplus)}{D_L}\,.
\label{eq:beta}
\end{equation}

Neglecting the acceleration in the Earth's motion and the
line-of-sight motion of the source and lens permits a simplification
of $\bm{\beta}$ to a linear function of time, though here we will
make use of the more general expression in equation (\ref{eq:beta}).

For a point-mass lens the observed position on the sky,
$\bm{\theta}$ is given by the thin-lens equation.  Usually this is
expressed as a potentially multi-valued function of the projected
source position.  However, we need only to identify the projected
source position in terms of the observed position $\bm{\theta}$,
given by
\begin{equation}
\bm{\beta} = \bm{\theta} \left(1 - \frac{\theta_E^2}{\left|\bm{\theta}\right|^2} \right)\,.
\end{equation}
Since surface brightness is conserved, this then immediately defines
the lensed image in terms of the intrinsic
\begin{equation}
I_{\rm lens}(\bm{\theta}) = I_{\rm int}\left[\bm{\beta}\left(\bm{\theta}\right)\right]\,.
\end{equation}

Finally, we convolve the lensed image with a realistic, if not
pessimistic, beam.  For this we assume a 43~GHz beam,
$B(\bm{\theta})$ is an anisotropic Gaussian, consistent with the
beam in \citet{2011A&A...525A..76L}: semi-minor axis of $0.5$~mas and semi-major axis of
$1.4$~mas, oriented $12^\circ$ East of North.  The large aspect ratio
is a result of a combination of the low declination of the Galactic
center and the North American location of the VLBA antennae.  In
practice, this resolution is lower than may be achieved by the VLBA at
43~GHz; in \citet{2003ApJ...587..208R} the interstellar scatter broadening limited the
size of the array that could be effectively employed.  Outside of the
Galactic center it should be possible to increase the resolution by a
factor of $2$--$3$, though we adopt the \citet{2011A&A...525A..76L} beam.
The resulting observed intensity distribution is then
\begin{equation}
I_{\rm obs}(\bm{\theta}) = \int d^2\!\theta'
B(\bm{\theta}-\bm{\theta}')
I_{\rm lens}(\bm{\theta}')\,,
\end{equation}
where prior to the convolution the angular positions are converted
into equatorial coordinates.

All that remains is to specify the relative positions, velocities, and
mass of the source and lens.  We do this for a handful of illustrative
examples in the following sections.

\subsection{Example Lensed Maser Images}

As fiducial source parameters we assume $D_S=8$~kpc and a proper
velocity of $120~{\rm km~s^{-1}}$, consistent with sources in
Galactic bulge.  The fiducial lens parameters are $D_L=4$~kpc and a
proper velocity of $30~{\rm km~s^{-1}}$, consistent with the velocity
dispersion in the Galactic disk.  The relative positions of the source
and lens are chosen so that minimum impact parameter is $0.3~$mas
oriented in declination.  These result in typical images that are
illustrative of the general situation.  All that remains is to specify
the mass of the lens.

\begin{figure*}
\begin{center}
\includegraphics[width=\textwidth]{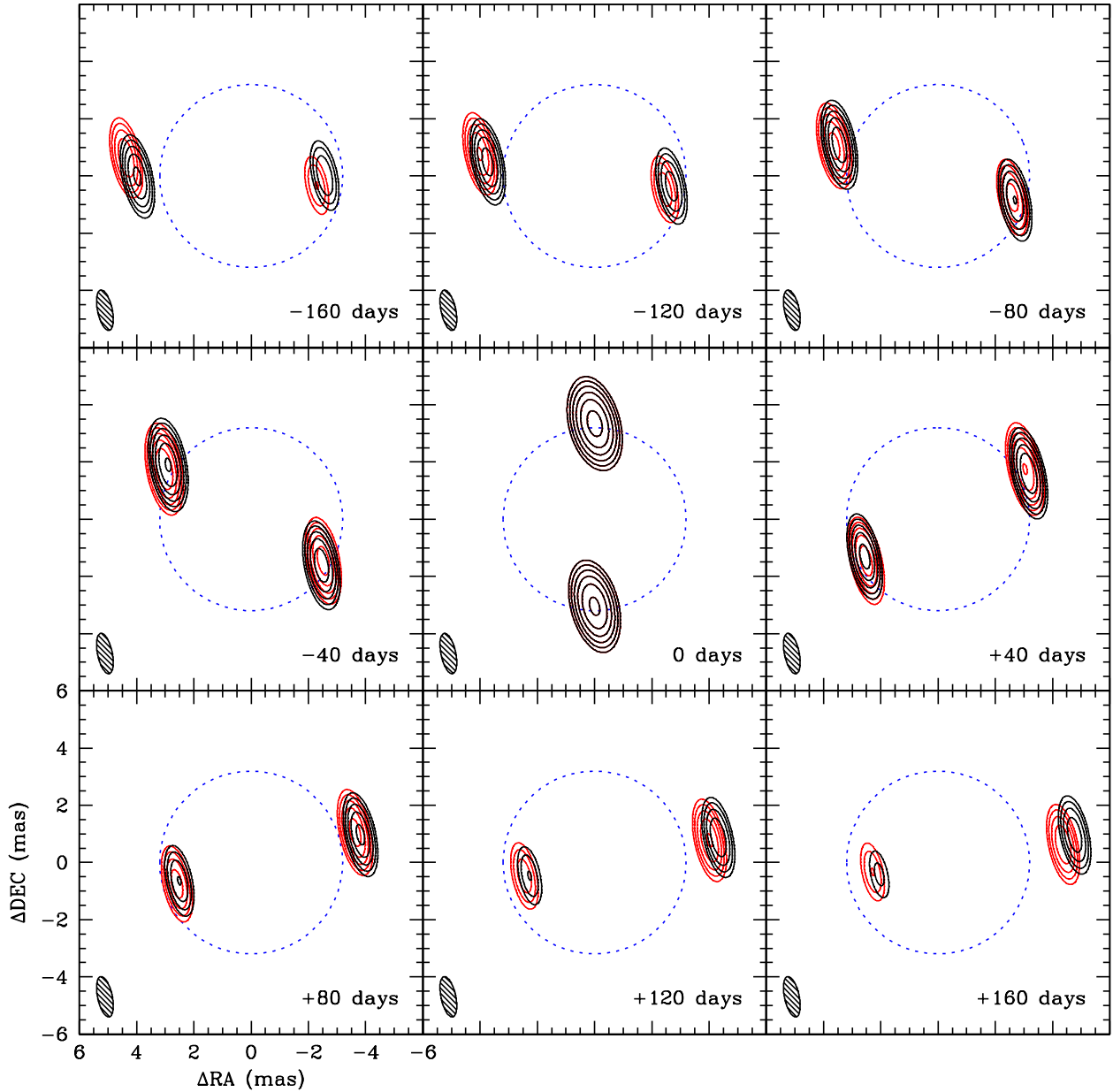}
\end{center}
\caption{Example $10~M_\odot$ black hole microlensing event.  Impact parameter of
  0.3~mas, relative lens-source velocity of $150~\kms$, 1~AU FWHM maser,
  and radio beam typical of Galactic center observations with the VLBA
  (semi-minor/major axes of 0.5~mas and 1.4~mas, respectively, with a
  position angle of $12^\circ$ east of north).  Red contours show
  images ignoring the orbital motion of the Earth, while black
  contours include the parallax.  For reference, the blue dotted line
  shows the Einstein ring.} \label{fig:bhi}
\end{figure*}

The first case we consider is that of a $10M_\odot$ black hole, for
which $\theta_E=3.2$~mas.  A time-sequence of images for our fiducial
case is shown in Figure \ref{fig:bhi}, covering nearly a year.  Since
the minimum impact parameter is small in comparison to $\theta_E$,
multiple images are present.  Because $\theta_E$ is also considerably
larger than the intrinsic source and beam sizes, these are well
resolved, suggesting that detecting and interpreting the features of
strong lensing by black holes will be straightforward.  In particular,
a direct measurement of $\theta_E$ is possible simply by fitting the
spot separations, though we defer to how well this may be done in
practice to Section \ref{sec:pe}.

\begin{figure*}
\begin{center}
\includegraphics[width=\textwidth]{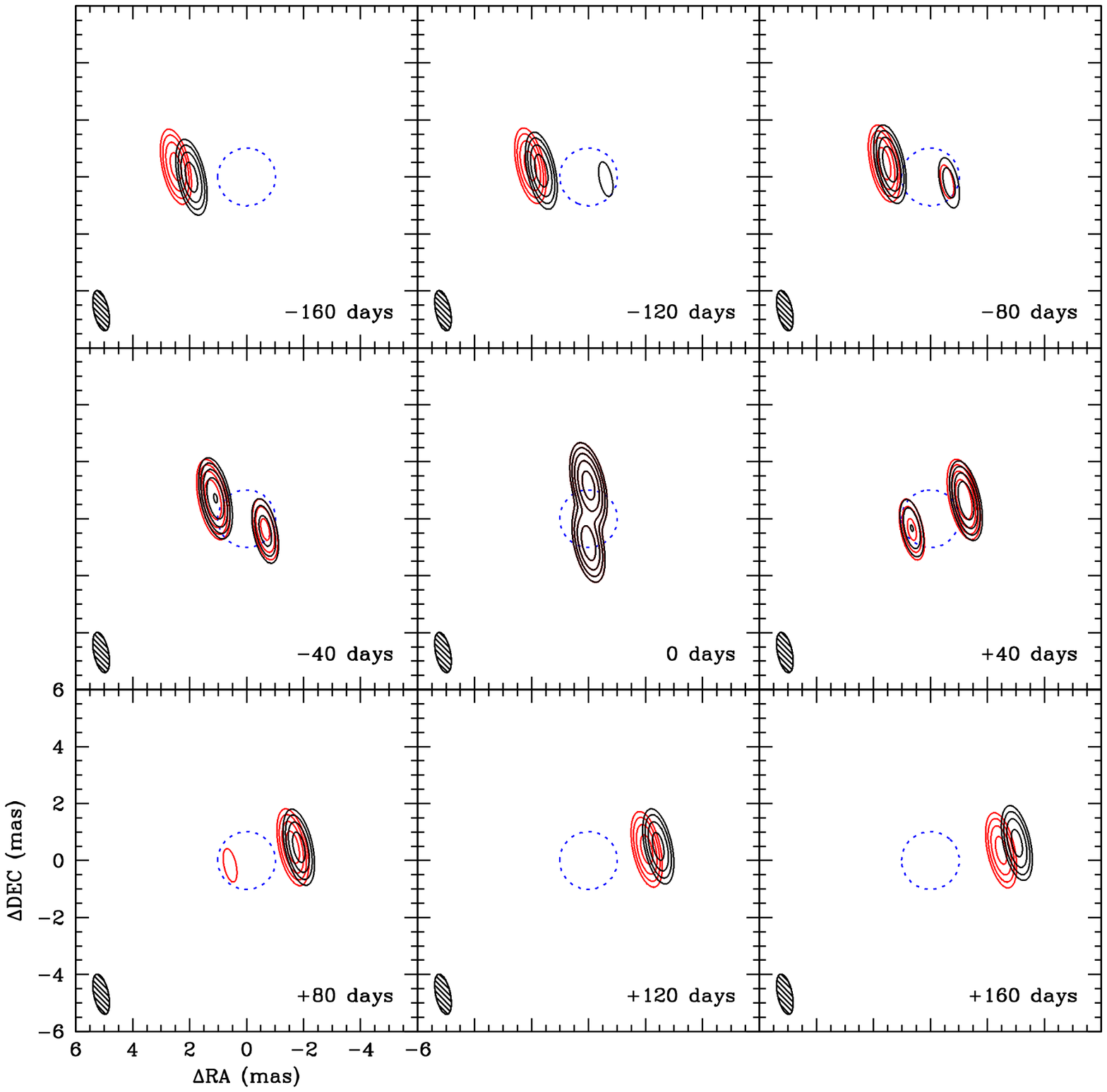}
\end{center}
\caption{Example $1~M_\odot$ stellar microlensing event.  Impact parameter of
  0.3~mas, relative lens-source velocity of $150~\kms$, 1~AU FWHM maser,
  and radio beam typical of Galactic center observations with the VLBA
  (semi-minor/major axes of 0.5~mas and 1.4~mas, respectively, with a
  position angle of $12^\circ$ east of north).  Red contours show
  images ignoring the orbital motion of the Earth, while black
  contours include the parallax.  For reference, the blue dotted line
  shows the Einstein ring.} \label{fig:wdi}
\end{figure*}

The same qualitative conclusions hold for a $1M_\odot$ lens,
indicative of a white dwarf, shown in Figure
\ref{fig:wdi}.\footnote{Recall that a solar-type star 
  would not modify the color of the combined lens-source system
  sufficiently to place it outside of the allowed region in the CMD.}
Despite the smaller Einstein angle ($\theta_E=1.0$~mas), multiple
images are resolved, again implying that $\theta_E$ can be directly
measured.  Where these differ most dramatically is near peak
magnification (center panel in Figures \ref{fig:bhi} and
\ref{fig:wdi}); the more massive lens produces correspondingly larger
distortions in the image as a result of the better resolution of the
Einstein ring.  The microlensing event is also significantly shorter.

The effect of the accelerated observer frame, due to the Earth's
orbital motion, is evident in the asymmetric entrance and exist from
the microlensing event.  However, this effect may be confused with
features arising from the impact of the asymmetric beam and arbitrary
event orientation.  Thus in Figures \ref{fig:bhi} and \ref{fig:wdi} we
have also shown by the red contours the images with the Earth's
orbital motion neglected.  From these it is clear that asymmetry in
the temporal evolution of the image structure at early and late times
is a robust indicator of parallax.

\subsection{Parameter Estimation} \label{sec:pe}

The primary parameters of interest are those of the lens: mass,
velocity, distance.  Given only the magnification light curve these
are degenerate, constrained only by the event time-scale and peak
magnification.\footnote{This may be improved substantially through the
  exploitation of the impact of parallax on the light curve \citep{gould92,rah03}.}
As a consequence, all that can be recovered even in principle are the
projected impact parameter in unites of $\theta_E$, $t_E$, and the
time of maximum magnification.
However, images like those in Figures \ref{fig:bhi} and \ref{fig:wdi}
introduce at least two key additional observables: $\theta_E$ and a
lensing parallax.

The peak magnification occurs at the projected closest approach, for
which the projected impact parameter is $\beta_{\rm min}$ and the
magnification is
\begin{equation}
A_{\rm max} = \frac{\beta_{\rm min}^2 + 
  2\theta_E^2}{\beta_{\rm min}\sqrt{\beta_{\rm min}^2+4\theta_E^2}}\,.
\end{equation}
The degree to which $A_{\rm max}$ may be measured depends upon the
quality of the photometry that can be performed, and thus is likely to
be limited by systematic errors in the radio flux calibration 
(flux calibration accuracy for the VLBA and the VLA at 43 GHz would be roughly $10\%$).
At this time the separation between the centroid of the
primary and secondary images is
\begin{equation}
\Delta\theta = \sqrt{\beta_{\rm min}^2 + 4\theta_E^2}\,,
\end{equation}
which may be measured directly from the images to a precision that
exceeds the interferometric beam width by up to an order of magnitude
for high signal-to-noise detections \citep{2003ApJ...587..208R}.  Thus, together,
these yield a high-precision estimate of $\theta_E$:
\begin{equation}
\begin{aligned}
\theta_E 
&=
\frac{\Delta\theta}{\sqrt{2}}\left[A_{\rm max}\sqrt{A_{\rm max}^2-1}-(A_{\rm max}^2-1)
\right]^{1/2}\\
&\approx
\frac{\Delta\theta}{2}\left(1 - \frac{1}{8 A_{\rm max}^2}\right)\,,
\end{aligned}
\label{thetaE}
\end{equation}
where the latter expression is for large $A_{\rm max}$, and a
fractional precision of 
\begin{equation}
\frac{\sigma_{\theta_E}^2}{\theta_E^2} 
\approx
\frac{\sigma_{\Delta\theta}^2}{\Delta\theta^2}
+
\frac{\sigma_{A_{\rm max}}^2}{4 A_{\rm max}^6}\,.
\end{equation}
When the uncertainty is dominated by the astrometric uncertainty, this
implies that for a $10M_\odot$ black hole $\theta_E$ can be measured
to within $\approx1\%$.

Even without detailed event modeling, $\theta_E$ immediately enables
the determination of the relative proper motion from
equation (\ref{eq:tE}):
\begin{equation}
\bm{\mu}_L-\bm{\mu}_S = \frac{\theta_E}{t_E} \bm{\hat{u}}\,,
\end{equation}
where $\bm{\hat{u}}$ is the asymptotic direction of the vector
between the first and second images.
The source proper motions can be directly measured by follow-up
observations; the angular velocities of individual maser spots have
been measured with an accuracy better than $1~{\rm mas~yr^{-1}}$.
As with the source positions, this is limited by the electron
scattering \citet{2003ApJ...587..208R}, and thus for sources beyond the central pc it should be
possible to measure this nearly an order of magnitude better, giving a
typical accuracy of $0.1~{\rm mas~yr^{-1}}$, corresponding to a
physical velocity of $4~{\rm km~s^{-1}}$ at the distance of the
Galactic center.  With a typical lens apparent velocity of 
$30~{\rm km~s^{-1}}$, comparable to the velocity dispersion within the
disk, this induces to a roughly 10\% uncertainty in the inferred
$\bm{\omega}_L$.

The remaining degeneracy with $D_L$ may be broken via the measurement
of a parallax from the microlensing event itself.  The sources of most
interest also are expected to have the longest $t_E$, and thus
encompass a substantial fraction of a year.  As made explicit in
Figures \ref{fig:bhi} and \ref{fig:wdi}, long $t_E$ permit
measurements of the impact of the Earth's orbital motion, i.e., a
parallax.  This comes in two forms.

\begin{figure}
\begin{center}
\includegraphics[width=\columnwidth]{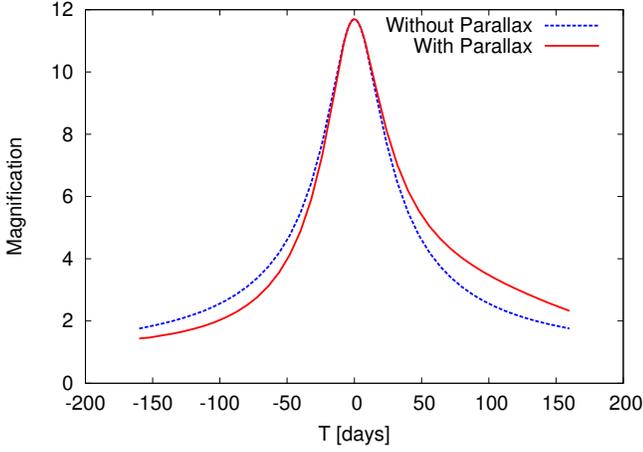}
\end{center}
\caption{Example light curve for the microlensing event shown in
  Figure \ref{fig:bhi}.  The blue dashed and solid red lines correspond to when
  the effect of parallax is neglected and included, respectively.  On
  $\pm50$~day time-scales the impact of parallax is clearly visible,
  resulting in roughly 30\% asymmetries in the light curve.} 
\label{fig:lightCurveBH} 
\end{figure}

First, the light curve itself is asymmetric as a result of the
modified evolution of the Earth-lens-source alignment \citep{2003MNRAS.339..925S}. 
Figure \ref{fig:lightCurveBH} shows the effect of parallax on 
light curve of a $10~M_{\odot}$ black hole lensing event. Since the
Earth accelerates substantially during this period, the additional
component is distinguishable from the otherwise unknown but
essentially fixed velocities of the source and lens, resulting in an
asymmetry in the magnification light curve.  While the maser emission
and optical stellar emission are not spatially coincident, the offset
in the distance is much smaller than $D_S$, and thus $D_L$ can be
reconstructed from either the optical or infrared light curves.  Typical 
uncertainties of $\%10$ in radio flux calibration may preclude the use 
of radio light curve for this purpose. Since
Miras are among the brightest stars in the microlensing survey fields,
they may be good candidates for obtaining $D_L$ in this way, assuming
the underlying variability can be adequately modeled.

\begin{figure}
\begin{center}
\includegraphics[width=\columnwidth]{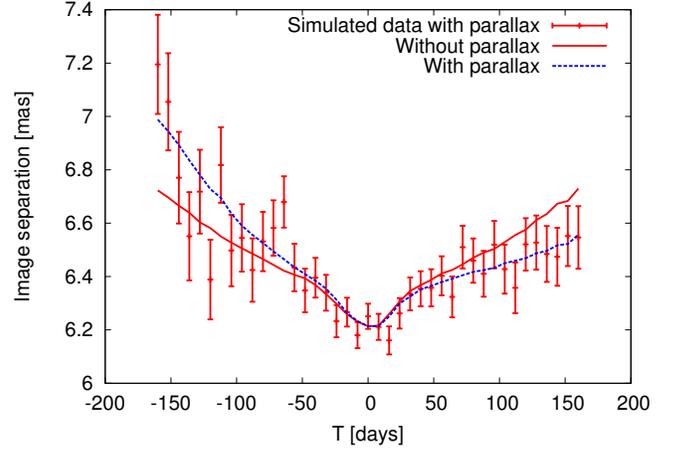}
\end{center}
\caption{Evolution of the image separations for the microlensing event
  shown in Figure \ref{fig:bhi}.  Centroid errors are constructed
  assuming a flux limit of $10\%$ the maximum flux.
  The dashed and dotted lines correspond to the
  expected image separations when the effect of parallax is
  neglected and included, respectively.  As with the light curve in
  Figure \ref{fig:lightCurveBH}, the largest impact of parallax occurs
  at early and late times, causing deviations of roughly 50\% of a
  beam width on year time-scales.
} \label{fig:bhseparation}
\end{figure}

Second, as described in the previous section, the underlying impact on
the lensed images themselves is directly measurable.  Note that this
is essentially the same effect, the asymmetry in the magnification
light curve arises from the asymmetry in the evolving image
structure.  However, unlike the parallax effect in the light curve, the image
structure is insensitive to the Mira variability.  While it is
possible to attempt a full fit to the sequence of images, most of the
information is contained in the locations of the multiple images.
Thus, shown in Figure \ref{fig:bhseparation} is the angular separation of image components
as a function of time for the sequence of images in Figure
\ref{fig:bhi}.

The mock data shown in figure \ref{fig:bhseparation} provide a convenient way to estimate the accuracy with which the parallax parameter can be reconstructed.  We compute the centroid positions of the multiple image components by performing a maximum-likelihood fit of the positions of beam-convolved point sources.  The uncertainty in the centroid is constructed assuming a flux limit of 10\% of the maximum brightness of an unlensed maser spot, and introduce corresponding Gaussian fluctuations in the mock data (see in Fig. \ref{fig:bhseparation}).  At late times the uncertainties grow as a result of the dimming of one of the two images far from peak magnification.

As a simple model we write the impact parameter as a linear combination of impact parameters with and without parallax effect included:
\begin{equation}
\beta = \lambda\beta_{p}+(1 - \lambda)\beta_{np} \,,
\label{eq:betalambda}
\end{equation}
where $\beta_{np}$ and $\beta_{p}$ are the angular separations when the orbital acceleration of the Earth is ignored and included assuming $D_L=D_S/2$.  As shown in Appendix \ref{app:PEMI}, the interpolation parameter is related to the lens distance by $\lambda=(D_S-D_L)/D_L$, and hence a measurement of $\lambda$ corresponds to a measurement of the lens position.  That is, keeping all angular measurements fixed, fitting the evolving image separation directly probes the lens distance.

A maximum-likelihood fit recovers $\lambda = 0.84\pm0.20$, where the 1$\sigma$ errors are indicated.  That is, for typical parameters the impact of parallax can be detected within the radio images at more that 4$\sigma$.  For a given $\lambda$ the corresponding estimates of the lens distance and its uncertainty are
\begin{equation}
  D_L = \frac{D_S}{1+\lambda}
  \quad\text{and}\quad
  \frac{\sigma_{D_L}^2}{D_L^2} = \frac{\sigma_\lambda^2}{(1+\lambda)^2}\,,
\end{equation}
and thus, we recover $D_L = 4.3\pm0.5~{\rm kpc}$.  That is, typically, $D_L$ can be reconstructed with an accuracy of roughly 10\%.

With an estimate of the lens distance it is possible to reconstruct the lens mass and transverse velocity from $\theta_E$ and $\bm{\omega}_L$.  The mass may be estimated given measurements of $\theta_E$ and $\lambda$ via
\begin{equation}
M = \frac{c^2}{4 G} \theta_E^2 \frac{D_S}{\lambda}\,.
\end{equation}
Assuming the uncertainty in $\theta_E$ is negligible and in $D_S$ is of order 10\%, the fractional uncertainty in $M$ is given by that in $\lambda$, and thus the mass can be estimated with an accuracy of roughly 14\%.
The lens transverse velocity is given by
\begin{equation}
\bm{v}_L = D_L \bm{\omega}_L\,.
\end{equation}
Assuming the typical uncertainties of 10\% in both quantities, the transverse velocity can be typically reconstructed to 14\% as well.

\subsection{Binaries}

\begin{figure*}
\begin{center}
\includegraphics[width=\textwidth]{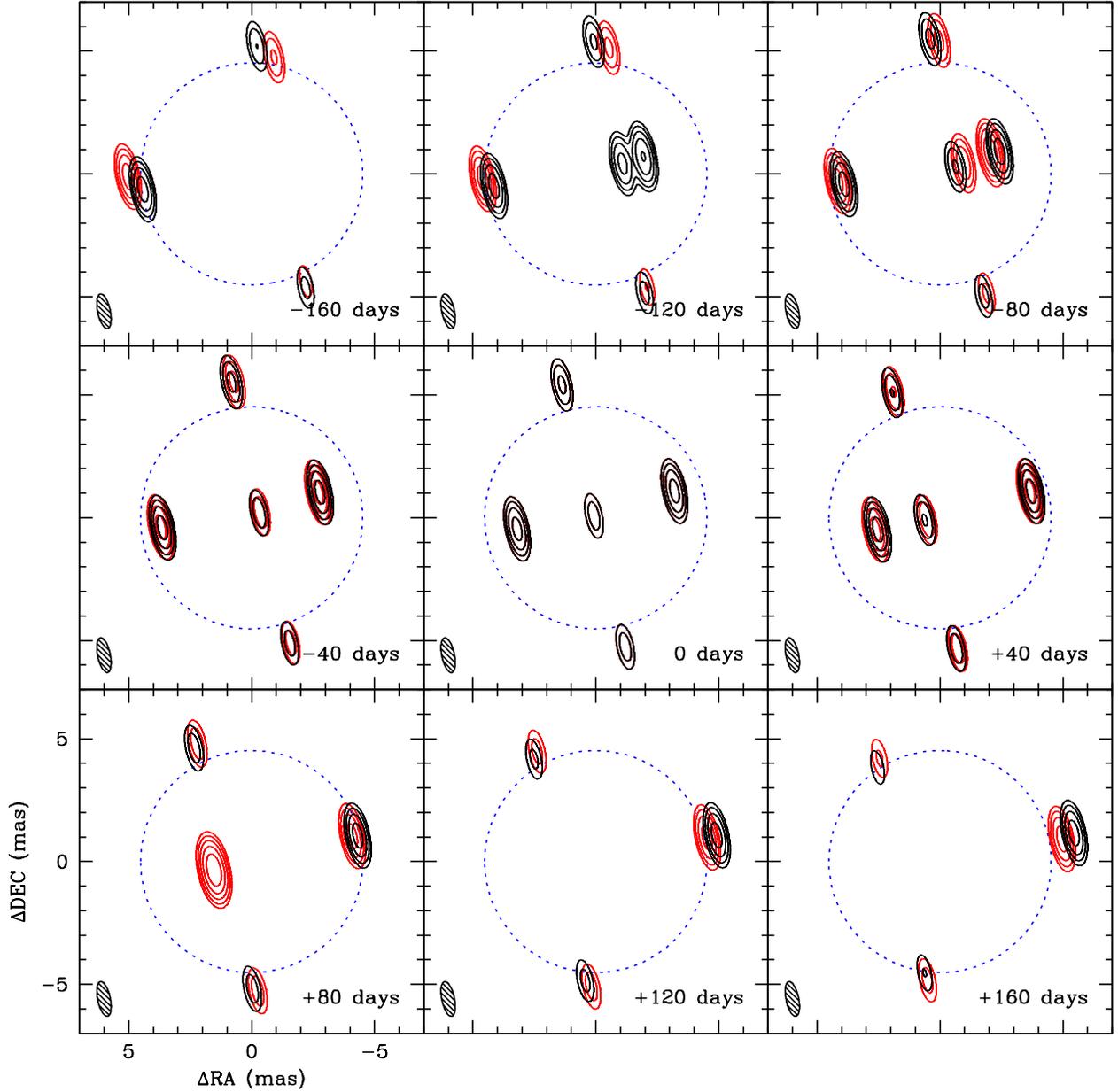}
\end{center}
\caption{Microlensing event example for a black hole-black hole binary
  system of equal mass ( $10~M_\odot$ each ) with orbital period of 30~yr.  Impact parameter of 0.3~mas, relative lens-source velocity
  of $150~\kms$, 1~AU FWHM maser, and radio beam typical of Galactic
  center observations with the VLBA (semi-minor/major axes of 0.5~mas
  and 1.4~mas, respectively, with a position angle of $12^\circ$ east
  of north). Red contours show images ignoring the orbital motion of
  the Earth, while black contours include the parallax.  For
  reference, the blue dotted line shows the Einstein ring.}
\label{fig:BinLns1}
\end{figure*}

Resolving the lensed images provides a natural way to distinguish
between single and double lenses, i.e., isolated and binary black
holes.  The latter, black hole-black hole binary systems are of particular interest
as a potential gravitational wave source \citep{LIGO02}. At the same time they have 
the potential to both inform and leverage the observations of X-ray binaries \citep[see, e.g.,][]{2006ARA&A..44...49R}.  Figure \ref{fig:BinLns1} shows a typical binary image,
which may be immediately distinguished from the Figure \ref{fig:bhi}
by both the complicated lensed-image morphology and, more directly,
the presence of a third image.

The efficiency with which binaries can be detected depends strongly on
their projected angular separation.  Sufficiently compact binaries
will appear as a single lens, while the components of sufficiently
wide binaries will produce independent lensing events.  After
conducting a suite of numerical experiments with binaries in different
orientations, peculiar velocities, and impact parameters, we have
found that binaries with projected angular separations spanning an
order of magnitude about the Einstein angle can be readily
identified.  The case shown in Figure \ref{fig:BinLns1} has a binary
angular separation comparable to the individual Einstein radii of the
black holes, $3.2~\rm mas$.  

As with the isolated case, the motion of the Earth induces shifts in
the image locations due to parallax, and thus the lens distance can
in principle be extracted.  Unlike the isolated case,
typically the image morphology during maximum magnification is
complicated. This is mitigated by the fact that the parallax signal is largest when the lens-source separation is largest, and thus where the binary will appear most point-like.  Nevertheless, we leave binary parameter estimation for future work.

\section{Discussion}\label{sec:disc}

With moderate changes in strategy existing and future optical/infrared
microlensing surveys should achieve black hole event rates of roughly $0.1~{\rm yr}^{-1}$. 
Dedicated radio microlensing surveys could reach event rates an order of
magnitude higher.  As a result, radio imaging of microlensing events can
provide the novel ability to produce a large sample of stellar remnants with
known masses, positions (including distances), and velocities.  Importantly,
this sample would not suffer from the standard biases that plague studies
based on binaries; the selection effects associated with microlensing
surveys are well understood and can be modeled and removed.  This enables
a variety of probes of astrophysical phenomena, which we discuss here.

\subsection{Massive Star Evolution}

The large event rates enables the systematic
construction of a black hole mass function over a decade.  For radio
microlensing surveys (see below) this rate can be an order of
magnitude larger.  Unlike X-ray binaries, for microlensing the
observing biases are well understood and can be addressed via
direct modeling, resulting in an accurate representation of the mass
distribution of stellar mass black holes in the Milky Way.

The mass distribution of black holes would be immediately diagnostic
of the dynamics of core collapse supernovae, the cataclysmic events
surrounding their formation.  Typically, accretion of fallback
material produces a mass gap between neutron stars and black holes, a
region of relative paucity in the black hole mass function extending
to masses well above $2M_\odot$
\citep{1996ApJ...457..834T,2002RvMP...74.1015W,2008ApJ...679..639Z}
. In all cases, key uncertainties in the stellar evolution (e.g.,
wind loss rates, supernovae energetics, etc.) impact the final mass
distributions \citep{gallart05}. There have been studies on the effect of variations in neutrino 
mechanism for core collapse supernovae on the theoretical black hole mass function 
\citep{2015ApJ...801...90P} and using the observed black hole mass function there already have been   
constraints on core collapse supernova \citep{2015MNRAS.446.1213K}.  
For compact binary systems, the location and depth of the mass gap is
further sensitive to the binary evolution history
\citep{2002ApJ...578..335F}.
While tentative evidence for a mass gap has already been reported based on
the inferred masses of black holes in X-ray binaries
\citep{Ozel10}, this is necessarily subject to
the variety of strong selection effects and substantial uncertainties
mentioned above.
Moreover, studies of X-ray binaries are fundamentally limited by the
small number of binaries currently known.

In contrast, radio-imaged microlensing events provide a means to
systematically accrue large, debiased samples of Galactic black holes,
limited only by survey duration.  Thus, it offers both a method to
assess the poorly known biases inherent in X-ray binaries and
ultimately to directly access key elements of massive star evolution.

Furthermore, radio-imaged microlensing events will produce estimates of the
lens positions and tangential velocities.  Thus, together with the
mass function it is possible in principle to construct a Galactic
black hole distribution function.  The velocity distribution will
necessarily be strongly impacted by the dynamics of core collapse
supernovae via supernova kicks \citep{Willems05}, as well as the subsequent
evolution of the Galactic remnant population.\footnote{While the
  dynamical relaxation time of   the Galaxy is of order 
  $10^4~{\rm Gyr}$, black holes are much more  massive than the
  typical Galactic object, and thus can have relaxed substantially.}

Galactic black holes also serve as a fossil record of massive star
formation.  Therefore, detailed comparisons of the
local\footnote{Necessarily accounting for potentially significant 
differences in relaxation rates.} densities of stellar mass black
holes, white dwarfs, and sub-solar stars yield an estimate of the
potentially temporally and spatially varying stellar initial mass
function.  In particular, this provides a unique ability to constrain
the history of the high-mass end of the initial mass function.

\subsection{Gravitational Wave Experiments}

The measurement of the mass of a single isolated stellar-mass black
hole would provide an important calibration for population synthesis
computations, performed primarily to produce rate estimates for
gravitational wave detectors \citep{belczynski02}.  This is the consequence of the
typical nature of lenses in contrast to the particular
evolutionary history of X-ray binaries, currently used to calibrate
the variety of uncertainties in the rate estimates.  

More generally, the binarity rate and source distribution are key
inputs into black hole binary gravitational wave source population
estimates.
High-mass X-ray binaries, the objects likely progenitors of the black
hole-black hole binaries visible by experiments like LIGO \citep{abadie10}, are
necessarily young by virtue of the large mass of the secondary,
providing a strong bias towards populations associated with the recent
star formation history of the Milky Way.  In stark contrast, binary
mergers are expected to occur very long times after their formation,
and therefore indicative of the integrated star formation history of
the Milky Way.
The detection of a black hole-black hole binary presents a
means to study the binary black hole population over the entirety of
the Galactic history.  Unfortunately, the lower limit on the projected
binary separation for which radio-imaged microlensing events can
efficiently detect binarity (roughly 1~mas at 8~kpc) corresponds to orbital
periods of years, and thus will not present candidate gravitational
wave sources.  Nevertheless, it would provide a significant calibration
of binary formation rates, and therefore black hole-black hole binary
populations.

\subsection{Isolated Neutron Stars}

Thus far we have focused primarily on black hole lenses.  Nevertheless,
it is possible to directly detect solar mass objects as well.  While we
have included a discussion of the white dwarf lensing rates we have
largely ignored lensing events from isolated neutron stars.  An estimate
of the neutron star lensing rate is nevertheless possible using the black
hole lensing rate.  The two key assumptions are that the massive stellar
progenitors of black holes and neutron stars follow a Salpeter initial mass
function and that the neutron stars receive kicks during formations.

From the first we estimate the ratio of the total number of Galactic neutron
stars to the total number of Galactic black holes:
\begin{equation}
  \frac{N_{\rm NS}}{N_{\rm BH}}
  \approx
  \left(\frac{M_{\rm ZNS}}{M_{\rm ZBH}}\right)^{-1.35}
  \approx
  4\,,
\end{equation}
where $M_{\rm ZNS}\approx8\,M_\odot$ and $M_{\rm ZBH}\approx21\,M_\odot$
are the minimum zero-age main-sequence masses of stars that form neutron
stars and black holes, respectively \citep{2002RvMP...74.1015W}.

The second assumption modifies the volume occupied by neutron stars.  Stars formed
at an initial radius $r_{\rm init}$ with a natal kick ($v_{\rm kick}$) larger
than the circular velocity ($v_{\rm circ}$) will isotropize within a radius of
\begin{equation}
  r_{\rm max} \approx r_{\rm init} e^{v_{\rm kick}^2/2 v_{\rm circ}^2}
  \approx 7 r_{\rm init}\,,
\end{equation}
where we have used typical values $v_{\rm kick}\approx400~\kms$ and
$v_{\rm circ}\approx200~\kms$.  Thus, neutron stars formed in the Galactic
bulge will be distributed across a volume nearly 400 times larger than similarly
formed black holes, and thus exhibit a number density 400 times smaller.
The net result is that for typical numbers the rate of neutron star microlensing
events is expected to be roughly 1\% of that for the black holes, justifying our
neglect of neutron stars.

However, we caution that this conclusion is extremely sensitive to the typical
kick velocities.  If the typical kick velocity is $300~\kms$ the neutron
star lensing rate rises to 10\% of the black hole lensing rate.  More importantly,
if the neutron star kick velocity distribution contains a low-velocity tail or is
bimodal, as suggested by the large number of known Galactic neutron stars
\citep[see, e.g.,][]{1998ApJ...496..333F,2005ASPC..328..327P}, the bulge may
retain a large fraction of neutron stars originally formed within it.  In this
cases the neutron star and black hole lensing rates can be comparable.  Thus, the
relative frequency of neutron star and black hole events provides an additional
probe of the distribution of neutron star formation kicks.

If large numbers of neutron star lensing events are observed, radio-imaged
microlensing provides a novel way in which to directly measure the neutron star
mass function, independent of the biases inherent in the study of neutron star
binaries.  Unfortunately, given their large typical age ($\gtrsim1~{\rm Gyr}$)
these are all likely to be exceedingly dim, with luminosities of
$\approx10^{-11}$-$10^{-9}L_\odot$, and therefore not amenable to direct size
measurements, complicating any effort to directly constrain the high-density
nuclear equation of state.

Nominally, these objects would also fall below the death line for typical
pulsar magnetic field strengths ($10^{12}~{\rm G}$) and periods (1~s).  However,
due to magnetic field decay, implicated by the low surface fields in recycled
millisecond pulsars, these objects could still exhibit observable magnetospheric
emission; typical surface fields after a Gyr would then be expected to lie near
$10^{11}~{\rm G}$.  Nevertheless, many would still live near the death line,
and therefore would at best be transient radio sources.  Hence, in this case the
radio emission of isolated neutron stars found by radio-imaged microlensing would
provide an unbiased probe of neutron star magnetization and its evolution.

\section{Conclusions}\label{sec:C}

VLBI observations of radio-bright microlensing events affords the ability to resolve the multi-component structure of microlensing events at radio wavelengths.  Such images would necessarily break the degeneracies between distance, velocity, and source mass inherent in studies of the light curve alone by providing a direct measurement of the size of the Einstein ring and the presence of signatures of parallax in the evolving image structure and/or asymmetric light curves.  Thus, radio imaging enables the reconstruction of the lens parameters to better than 15\% accuracy.  The dominant source of this remaining uncertainty is due to the unknown distance and velocity of the lensed source, and thus these may be improved by roughly an order of magnitude by follow-up observations that constrain the source's proper motion and distance.  Furthermore, imaging provides an immediate method to detect binary lenses, albeit within a rather narrow range of orbital separations (corresponding to periods of roughly $2~{\rm yr}$--$20~{\rm yr}$.  It is noteworthy that this remains the case when the lens belongs to the otherwise unobservable Galactic remnant population, e.g., neutron stars and black holes.

Radio-bright microlensing events can be identified using existing optical/infrared microlensing surveys.  Mira variables exhibit SiO masers in their envelops and therefore provide a natural radio-luminous target.  Moderate modifications to existing survey strategies, some of which are already pursued by OGLE-IV, should produce events amenable to radio imaging at a rate of $\approx~2~{\rm yr^{-1}}$.  Of these, assuming the Galactic black hole population arises solely via the evolution of massive stars, roughly $0.1$ will due to black holes.  This rate will increase substantially if black holes compose a significant fraction of the Galactic dark matter budget.

Detections of Galactic black holes via microlensing provides an unbiased sample of massive star remnants.  The existing sample of stellar mass black holes, obtained via observations of X-ray binaries, necessarily suffers from strong, uncertain biases associated with the formation and evolution of tight massive-star binaries.  As a consequence, the detection and characterization of even a handful of black holes using radio-imaged microlensing events will inform the late-stage evolution of massive star, the energetics of their subsequent supernovae, and the event rates of current and future gravitational wave experiments.

Currently, the rate of radio-bright microlensing events is limited by the number of compact radio sources above the detection limits of existing VLBI facilities, and in particular the VLBA. Nevertheless, an order of magnitude increase in event rates can be obtained by improvements in the flux limits of VLBI observations, achieved, e.g., through increased  collecting area (as could be provided with long baselines in the next-generation VLA), expanded bandwidths and longer integration times, and dedicated radio-continuum microlensing surveys.    
Therefore, it is possible that in the near future it will be possible to double the number of known Galactic black holes with a survey extending two years, and begin the statistical study of the Galactic black hole properties within a decade.

\section*{Acknowledgements}
The authors would like to thank Andrew Gould, Bryan Gaensler, Geoffry Bower, Simone Portegeis Zwart, Leo Stein, and Niayesh Afshordi for helpful comments and discussions. We also thank the anonymous referee for many insightful comments that have significantly improved the accuracy and clarity of the presentation. 
A.E.B.~receives financial support from the Perimeter Institute for Theoretical Physics and the Natural Sciences and Engineering Research Council of Canada through a Discovery Grant.  Research at Perimeter Institute is supported by the Government of Canada through Industry Canada and by the Province of Ontario through the Ministry of Research and Innovation.

\bibliographystyle{mn2e} 
\bibliography{Bibliography}

\appendix
\label{appendix}

\section{Breaking degeneracy with combined observation}
In this section we introduce the mathematical basis of gravitational
microlensing, introduce the photometric and astrometric observations
of microlensing events and the way we can break the degeneracy between
the lens parameters.

\subsection{Lens Equation for Single Lens } \label{app:lens}
Single lens equation is given by geometrical relation between the
angular position of source $\beta$, angular position of the image
$\theta$ and Einstein angle $\theta_E$ as follows:
\begin{equation}
\theta^2 - \theta\beta - \theta_E^2 = 0,
\label{lequation}
\end{equation}
where $\beta$ and $\theta$ are given with respect the center of
coordinate system where lens is located. In the gravitational
microlensing $\beta$ changes with time on a straight line as
\begin{equation}
\beta^2=\beta_0^2 + \theta_E^2(\frac{t-t_0}{t_E})^2,
 \label{beta}
\end{equation}
where $\beta_0$ is the minimum impact factor and $t_0$ is the time of
closest angular approach of the lens and source. From the equation
(\ref{lequation}), we get two solutions for the position of two images
at two sides of the lens as follows:
\begin{equation}
\theta^\pm = \frac{\beta\pm\sqrt{\beta^2 + 4\theta_E^2}}{2}.
\label{solution}
\end{equation}
Substituting equation (\ref{beta}) in equation(\ref{solution}) results
in the radial dynamics of the images as a function of time. 

The angular distance between the two
images ($\Delta\theta = \theta^+ -\theta^-$), regardless of the
position of source, can be written in terms of the impact
parameter and the Einstein angle:
\begin{equation}
\Delta\theta = \sqrt{\beta^2 + 4 \theta_E^2} \label{dtheta}
\end{equation}
At the minimum impact parameter where $\beta = \beta_0$, two images
get nearest approach to each other and in this case the trajectory of
the source is perpendicular to the line connecting these two images

\begin{equation}
\beta_0 = \sqrt{\Delta\theta^2 - 4 \theta_E^2}. \label{beta0}
\end{equation}
We note that on the right hand side the term in the square-root should be
positive which implies $\Delta\theta\geq 2\theta_E$. So the separation
between the two images is always larger than the diameter of Einstein ring
while one image is inside and the other one is outside the Einstein
ring.

The other observable parameter which can be measured from the photometry
is the magnification of the source during the lensing.  Since in the
gravitational lensing the flux of light is conserved, the apparent change of area for lensed images
would lead to a change in the energy we receive from each image. The
area of images compared to the source size is given by the inverse of
determinant of Jacobian. Denoting the area for the two images by $A^+$
and $A^-$ the differential element of area of an image to that of the source is
given by:
\begin{equation}
A^\pm =
\frac{\theta^\pm}{\beta}\frac{\partial\theta^\pm}{\partial\beta}.
\label{dtdb}
\end{equation}
Substituting equation (\ref{solution}) in equation(\ref{dtdb}) the area for each images could be written as:
\begin{equation}
A^\pm =
\frac{1}{2}\left(\frac{\beta^2+2\theta_E^2}{\beta\sqrt{\beta^2+4\theta_E^2}}\pm
1\right)
\end{equation}
The overall magnification is the sum of these two terms,
\begin{equation}
A =| A^+| + |A^-|
=\frac{\beta^2+2\theta_E}{\beta\sqrt{\beta^2+4\theta_E^2}}.
\label{mag}
\end{equation}
Substituting equation for (\ref{beta}) the magnification is independent of
Einstein angle and only depends on $t_0$, $t_E$ and $u_0
= \beta_0/\theta_E$.

At the time of minimum approach the
magnification would have its maximum as a function of $\beta_0$ and $\theta_E$.  Substituting
$\beta_0$ from equation (\ref{beta0}) in (\ref{mag}),
maximum magnification can be obtained as a function of Einstein angle and angular
separation between the images as follows:
\begin{equation}
A_{\rm max} = \frac{\Delta\theta^2 -
  2\theta_E^2}{\Delta\theta\sqrt{\Delta\theta^2-4\theta_E^2}}
\end{equation}
As a result the Einstein angle can be directly extracted from $A_{\rm max}$ as described in equation (\ref{thetaE}).

If the distance of the source from earth is known (which is the case for
microlensing events in Galactic center)  measuring
$\theta_E$ can put constrains on the mass and the lens distance
from observer. On the other hand parallax measurements can impose
another constrain on the lens parameters and in theory can resolve
the lens parameters.

\subsection{Parallax Effect in Microlensing Light Curves}

The microlensing parallax effect results from the accelerating motion of the earth
around sun. It manifests itself as the relative motion of the source with respect
to lens deviates from a straight line. Assuming earth moves around
the sun in a circular orbit, the relative angular motion of the lens
and source with respect to the observer in the case of stationary source and lens is given by
\begin{eqnarray}
\pi_x &=& \pi\cos\xi(t),\\ \pi_y &=& \pi\cos\beta\sin\xi(t),
\end{eqnarray}
where
\begin{equation}
\pi =\frac{1}{\theta_E}(\frac{1 a.u.}{D_l}-\frac{1 a.u.}{D_s}),
\label{pi}
\end{equation}
$\beta$ is the angular deviation of the orbital plane with respect to
our line of sight, $\xi(t) = \omega t + \xi_0$, $\omega$ is the
orbital angular frequency and $\xi_0$ is the initial phase of the
earth orbit which can be set to zero at the equinox. On the other hand
relative displacement of the source with respect to position of the
sun on the lens plane is given by
\begin{eqnarray}
\beta_x^{S} &=& \beta(t)\cos\alpha -\beta_0(\cos\alpha+\sin\alpha),
\\ \beta_y^{S} &=& \beta(t)\sin\alpha+\beta_0(\cos\alpha-\sin\alpha),
\end{eqnarray}
where the superscript represents relative motion of the source with
respect to the sun and $\beta_0$ is the closest approach of the source
to the sun and $\alpha$ is the angle between the trajectory of source
and semi-major axis of the earth orbit.  Now combining these two
motions the relative motions of the source with respect to the earth
is given by
\begin{equation}
\vec{\beta}(t) = \vec{\beta^S}(t) - \vec{\pi}(t).
\end{equation}
The second term corresponds to a perturbation compared to the simple
microlensing events. The photometric effect of parallax
can be measured as an asymmetry of the microlensing light curve
\citep{firstparallax}. Also the astrometric effect of parallax is
observable, if the lensed images are resolved. By
combining equation (\ref{thetaE}) and (\ref{pi}) the
distance of lens from the observer ($D_l$) can be extracted. Substituting in the
definition of $\theta_E$, the mass of lens can be obtained. 
In fact the lens mass can be obtained only given $\theta_E$ and $\pi_r$
using the relation:
\begin{equation}
\frac{M}{M_\odot} = \frac{1}{8\pi_r} (\frac{\theta_E}{1~\text{mas}}).
\end{equation}

From photometric observations and microlensing light curves
the physical parameter that is obtained by fitting to the
light curve is the Einstein crossing time. Having measured this
parameter the transverse velocity of lens can be derived as $v_t =
D_l\theta_E/t_E$.

\subsection{Parallax Effect in Microlensing Images} \label{app:PEMI}

The acceleration of the Earth impacts the image separations for the same reason it affects the light curve.  Here we estimate the size of this effect, explicitly determining the way in which the lens distance enters.

We begin by defining averaged and perturbed lens and source positions,
\begin{equation}
  \bm{s} \equiv \bm{x}_S -  \bm{x}_\oplus = \bar{\bm{s}} - \bm{\delta}
  \quad\text{and}\quad
  \bm{\ell} \equiv \bm{x}_L -  \bm{x}_\oplus = \bar{\bm{\ell}} - \bm{\delta}\,,
\end{equation}
where $\bar{\bm{s}}$ and $\bar{\bm{\ell}}$ include the linear lens and source motions and the linear component of the Earth's motion.  The acceleration of the Earth is contained in $\bm{\delta}$ which is by construction quadratic in time to lowest order.  We will presume this to be small in comparison to $\bar{\bm{s}}$ and $\bar{\bm{\ell}}$.

Inserting these into equation (\ref{eq:beta}) and linearizing in $\bm{\delta}$ and ignoring terms of order $\beta^2$ yields
\begin{equation}
  \begin{aligned}
  \bm{\beta}
  &\equiv
  \left(\bm{1}-\frac{\bm{\ell}\bm{\ell}}{D_L^2}\right)\cdot\frac{\bm{s}}{D_S}\\
  &\approx
  \left(\bm{1}-\frac{\bar{\bm{\ell}}\bar{\bm{\ell}}}{D_L^2}\right)\cdot\frac{\bar{\bm{s}}}{D_S}\\
  &\qquad
  +
  \left( \frac{D_S}{D_L} - 1 \right)
  \left(\bm{1}-\frac{\bar{\bm{\ell}}\bar{\bm{\ell}}}{D_L^2}\right)\cdot\frac{\bm{\delta}}{D_S}\\
  &\qquad
  +
  \frac{\bar{\bm{\ell}}}{D_L} \frac{\bar{\bm{s}}}{D_L}\cdot\left(\bm{1}-\frac{\bar{\bm{\ell}}\bar{\bm{\ell}}}{D_L^2}\right)\cdot\frac{\bm{\delta}}{D_S}\,.\\
  &
  \end{aligned}
\end{equation}
Therefore, noting that $\bm{\beta}\cdot\bm{\ell}=0$, 
\begin{equation}
\beta \approx \beta_{np} + \left(\frac{D_S-D_L}{D_L}\right) \frac{\bm{\beta}_{np}}{\beta_{np}}\cdot\frac{\bm{\delta}}{D_S}\,,
\label{eq:betap}
\end{equation}
where $\bm{\beta}_{np}\equiv\left(\bm{1}-\bar{\bm{\ell}}\bar{\bm{\ell}}/D_L^2\right)\cdot\bar{\bm{s}}/D_S$ is the angular position of the lens ignoring the Earth's acceleration.  The corresponding angular separation between the multiple images is then given by equation (\ref{dtheta})
\begin{equation}
\Delta \theta \approx \Delta \theta_{np} + \left(\frac{D_S-D_L}{D_L}\right) \frac{\bm{\beta}_{np}}{\Delta\theta_{np}}\cdot\frac{\bm{\delta}}{D_S}\,.
\end{equation}\\
Note that equation (\ref{eq:betap}) interpolates between $\beta_{np}$ and $\beta$ as the lens distance changes.  With the definition
\begin{equation}
\beta_p \equiv \beta_{np} + \frac{\bm{\beta}_{np}}{\beta_{np}}\cdot\frac{\bm{\delta}}{D_S}\,,
\end{equation}
i.e., when $D_L=D_S/2$, we recover equation (\ref{eq:betalambda}),
\begin{equation}
  \beta = \lambda \beta_p + (1-\lambda) \beta_{np}\,,
  \tag{\ref{eq:betalambda}}
\end{equation}
where
\begin{equation}
\lambda \equiv \frac{D_S-D_L}{D_L}\,.
\end{equation}

\end{document}